\documentclass[prl,reprint,doublecolumn,superscriptaddress, nobalancelastpage,dvipsnames]{revtex4-2}

\usepackage[dvipsnames]{xcolor}

\usepackage{tikz}

\usepackage{etoolbox}
\patchcmd{\section}
  {\centering}
  {\raggedright}
  {}
  {}
\patchcmd{\subsection}
  {\centering}
  {\raggedright}
  {}
  {}
  
\usepackage{enumitem,kantlipsum}
\usepackage{tcolorbox}
\usepackage[sort&compress]{natbib}
\usepackage{soul}
\usepackage{physics}
\usepackage{graphicx}
\usepackage{epstopdf}
\usepackage{hhline}
\usepackage{verbatim}
\usepackage{float}
\usepackage{sidecap}
\sidecaptionvpos{figure}{c}
\usepackage{lipsum}
\usepackage{capt-of}
\usepackage{setspace}
\usepackage{dcolumn}
\usepackage{bm}
\usepackage{amssymb}
\usepackage{amsmath}
\usepackage{mathtools}
\usepackage{color}
\usepackage{multirow}
\usepackage{lipsum}
\usepackage[a4paper]{geometry}
\newgeometry{top=1.25cm,right=1cm,left=1.25cm,bottom=1.25cm}
\usepackage[colorlinks=true, urlcolor=blue, pdfborder={0 0 0}]{hyperref}
\hypersetup{
linkcolor=blue,
citecolor=blue
}
\DeclareMathOperator*{\argmin}{argmin}
\DeclareMathOperator*{\argmax}{argmax}

\draft%
\begin{document}

\title{\Large{\textbf{Benchmarking Adversarially Robust Quantum Machine Learning at Scale}}}

\author{Maxwell T. West} \email{westm2@student.unimelb.edu.au} \affiliation{School of Physics, The University of Melbourne, Parkville, 3010, VIC, Australia}
\author{Sarah M. Erfani}
\affiliation{School of Computing and Information Systems, Melbourne School of Engineering, The University of Melbourne, Parkville, 3010, VIC, Australia}
\author{Christopher Leckie}
\affiliation{School of Computing and Information Systems, Melbourne School of Engineering, The University of Melbourne, Parkville, 3010, VIC, Australia}
\author{Martin Sevior} \affiliation{School of Physics, The University of Melbourne, Parkville, 3010, VIC, Australia}
\author{\\ Lloyd C.L. Hollenberg} \affiliation{School of Physics, The University of Melbourne, Parkville, 3010, VIC, Australia}
\affiliation{ Center for Quantum Computation and Communication Technologies, The University of Melbourne, Parkville, 3010, VIC, Australia}
\author{Muhammad Usman} \email{musman@unimelb.edu.au}  \affiliation{School of Physics, The University of Melbourne, Parkville, 3010, VIC, Australia}
\affiliation{ Center for Quantum Computation and Communication Technologies, The University of Melbourne, Parkville, 3010, VIC, Australia}
\affiliation{Data61, CSIRO, Research Way Clayton, 3168, VIC, Australia}

\maketitle%

\onecolumngrid%

\noindent%
\textcolor{black}{\normalsize{\textbf{Machine learning (ML) methods such as artificial neural networks are rapidly becoming ubiquitous in modern science, technology and industry. Despite their accuracy and sophistication, neural networks can be easily fooled by carefully designed malicious inputs known as adversarial attacks. While such vulnerabilities remain a serious challenge for classical neural networks, the extent of their existence is not fully understood in the quantum ML setting. In this work, we benchmark the robustness of quantum ML networks, such as quantum variational classifiers (QVC), at scale by performing rigorous training for both simple and complex image datasets and through a variety of high-end adversarial attacks. Our results show that QVCs offer a notably enhanced robustness against classical adversarial attacks by learning features which are not detected by the classical neural networks, indicating a possible quantum advantage for ML tasks. Contrarily, and remarkably, the converse is not true, with attacks on quantum networks also capable of deceiving classical neural networks. By combining quantum and classical network outcomes, we propose a novel adversarial attack detection technology. Traditionally quantum advantage in ML systems has been sought through increased accuracy or algorithmic speed-up, but our work has revealed the potential for a new kind of quantum advantage through superior robustness of ML models, whose practical realisation will address serious security concerns and reliability issues of ML algorithms employed in a myriad of applications including autonomous vehicles, cybersecurity, and surveillance robotic systems.}}} \\ \\

\twocolumngrid%

\section{I. Introduction}

\noindent
The past decade has seen an extraordinary multidisciplinary uptake of machine learning (ML) methods, driven largely by the success of deep neural networks~\cite{lecun2015deep}, in a diverse set of scientific applications including, for example, image classification~\cite{he2016deep}, natural language processing~\cite{otter2020survey}, protein structure prediction~\cite{jumper2021highly}, and quantum circuit optimisation~\cite{fosel2021quantum}. At the same time, the rise of autonomous vehicles, drones and robots has seen ML technology increasingly enter industrial and military use, with attention naturally turning to the reliability of such ML tools in the face of malicious actors who may seek to exploit them~\cite{biggio2013evasion,szegedy2013intriguing,10.1145/2046684.2046692,kurakin2016adversarial,goldwasser2022planting}. Most notably, a serious blow to the prospect of reliably using neural networks in security conscious environments has been delivered in the discovery of adversarial attacks, and the resulting field of adversarial ML~\cite{note_2, biggio2013evasion,szegedy2013intriguing,10.1145/2046684.2046692,kurakin2016adversarial,goldwasser2022planting}. The key finding of adversarial ML is that even well trained, high performing neural networks will generally possess serious vulnerabilities to inputs which have been carefully crafted in order to deceive them, despite possibly being all but indistinguishable~\cite{szegedy2013intriguing} from genuine inputs which the network can classify easily. This discovery renders the use of artificial neural networks in environments where the input source cannot be trusted a dangerous proposition, and is a key roadblock on the path to widespread deployment of artificial intelligence in general scenarios~\cite{goldwasser2022planting}. Significant efforts have been made to strengthen neural networks against adversarial attacks~\cite{wong2020fast, madry2017towards, goodfellow2018making, miller2020adversarial, cohen2019certified, bai2021recent}, including for example the judicious injection of small amounts of randomness in order to disturb any sensitively constructed adversarial inputs~\cite{cohen2019certified}, or the inclusion of adversarial examples at training time in order to build robustness~\cite{wong2020fast, bai2021recent}, but a universal and reliable defence mechanism still remains elusive. While for now the long-term prospects of classical neural networks in the face of adversarial attacks remains unclear, recently attention has been turning to how emerging quantum ML solutions will fare against adversarial attacks. \\[4pt]
\begin{figure*}[ht]
 \begin{center}
 \includegraphics[width=\textwidth]{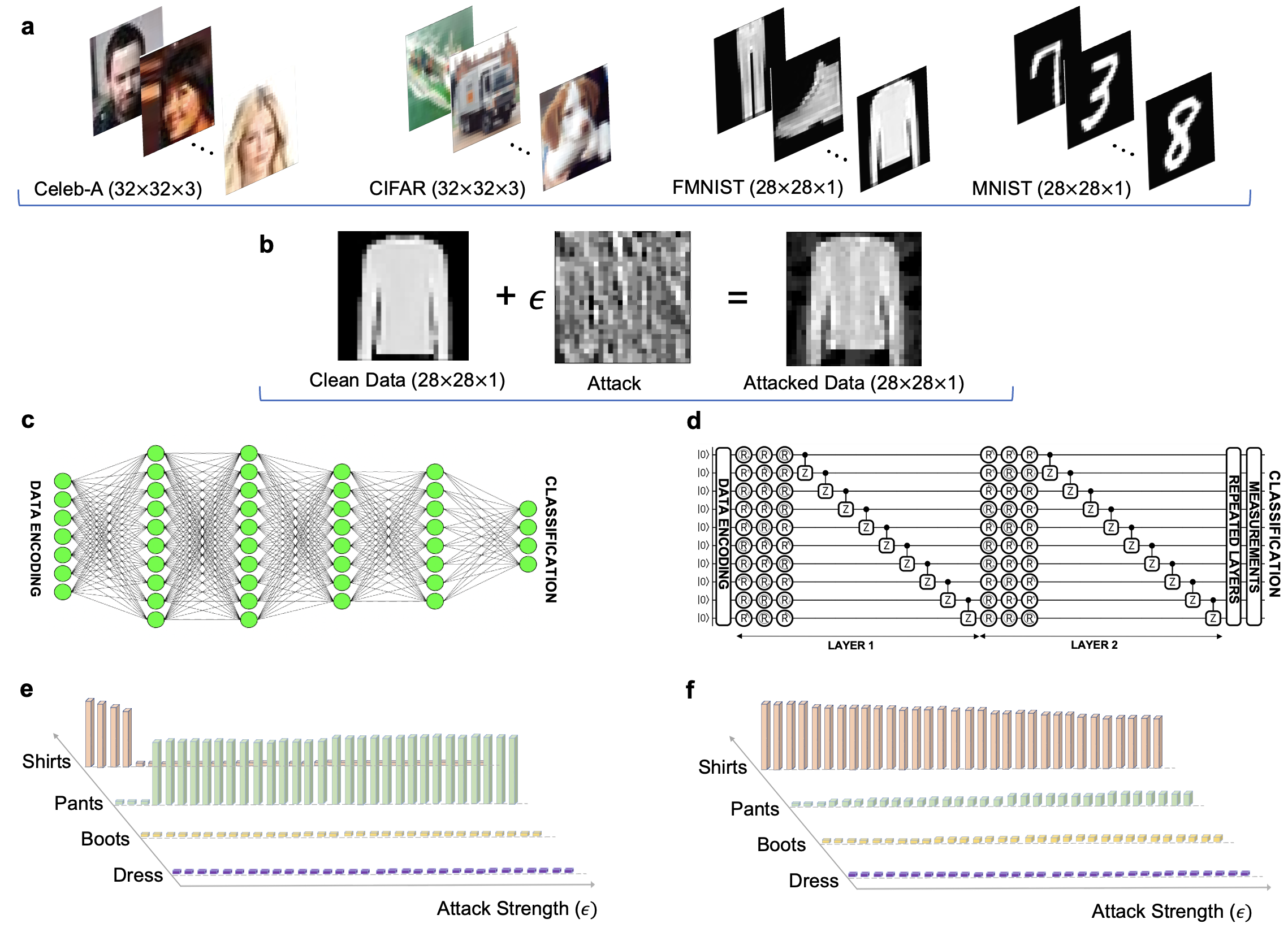}
  \caption{\label{fig:1}\textbf{Adversarial Machine Learning Benchmark Framework.}
A flowchart diagram of the quantum/classical adversarial ML framework developed to benchmark their robustness against a range of sophisticated adversarial attacks. \textbf{a.} A variety of image datasets (MNIST, FMNIST, CIFAR and Celeb-A) are selected to train and test classical and quantum ML networks. \textbf{b.} An example image from the FMNIST dataset along with an adversarial attack and the attacked image is shown. For each image dataset, adversarial attacks (PGD, FGSM, and AutoAttack) are generated with variable strengths controlled by $\epsilon$  (with respect to the $l_{\infty}$ norm). Importantly, we generate both quantum and classical attacks to test their transferability across networks, \textit{i.e.}, quantum attacks on classical networks and vice versa. \textbf{c.} Schematic diagram of a typical classical neural network is illustrated. In our work, two classical neural networks (ConvNet and ResNet) were trained to directly compare their performance against the quantum ML networks. The exact configurations of the classical networks are described in the Appendix. \textbf{d.} Schematic diagram of a quantum variational classifier (QVC) network trained and tested in this work. The QVC network consists of a data loading layer where the input image data is amplitude encoded into a quantum state. The variational part of the QVC consists of a repeated pattern of layers -- only two such layers are shown. We trained and tested a number of QVC networks with varying number of repeated layers, which are labelled based on the number of repeated layers, \textit{e.g.}, a QVC200 network contains 200 layers. The final stage of the network is a measurement layer which determines the classification label.  \textbf{e, f.} Here we depict the performance of a quantum and a classical network against an adversarial attack, where a small, carefully chosen perturbation is added to an image (as shown in \textbf{b}) which is then passed to both a neural network and a QVC for classification. The probabilities that are assigned to various labels by the networks are shown as a function of the strength of the attack. At a certain critical attack strength the correct label (``Shirts'', shown in orange) is no longer considered the most probable by the classical network (see \textbf{e}), and the attack has succeeded in fooling the model. Contrarily, the same attacked image when passed to a QVC is still correctly predicted as ``Shirts'' (see \textbf{f}), even when the attack strength is increased well beyond what is tolerated by the classical network.}
 \end{center}
 \end{figure*}
\noindent
Quantum ML is a new paradigm for the design of ML solutions~\cite{note_3, biamonte2017quantum} with the possibility of exploiting the capabilities of quantum computing for superior performance in ML applications, either by utilising quantum subroutines to enhance the performance of classical ML classifiers~\cite{hhl, childs2017quantum}, or through the development of classifiers which are themselves inherently quantum~\cite{beer2020training, havlivcek2019supervised, romero2017quantum, dallaire2018quantum, kehoe2021defence, heredge2021quantum, srikumar2022kernel}. Remarkable advances in both quantum hardware and software development \cite{jurcevic2021demonstration} have led to a great interest in developing and benchmarking a variety of quantum ML methods \cite{Coles2022qmlsurvey}. This naturally leads to an important question: Can quantum ML algorithms be designed to achieve superior defence against adversarial attacks compared to their classical counterparts? This is the key open question which we aim to address in this work. There have been a few recent studies which have explored the extent to which quantum classifiers themselves suffer from adversarial examples~\cite{lu2020quantum,quantum_cm,du2021quantum,guan2021robustness,weber2021optimal,ren2022experimental,liao2021robust,kehoe2021defence,west2022quantumising}, giving birth to a new field of quantum adversarial ML (QAML) \cite{West2022qamlsurvey}. However, the QAML literature has to date been limited to only small-scale proof-of-concept studies such as based on downscaled MNIST data \cite{kehoe2021defence, lu2020quantum} or other simple binary classification problems \cite{ren2022experimental}. Furthermore, the field has so far primarily focused on the vulnerability assessment for white box adversarial attacks, wherein the adversary is assumed to have complete knowledge of the target network \cite{quantum_cm, lu2020quantum, weber2021optimal, ren2022experimental}. By developing an adversarial ML benchmark framework (see Figure~\ref{fig:1} for a flowchart description), our work is the first to evaluate the robustness of quantum networks under true vulnerability tests through benchmarking their performance at scale, \textit{i.e.}, based on full-sized simple and complex image datasets without any downscaling and by creating both white and black box attacks. In contrast to white box attacks, a black box attack is generated without precise knowledge of the ML network structure, the phenomena which is also known as the transferability of adversarial examples in the ML literature~\cite{szegedy2013intriguing, ilyas2019adversarial, tsipras2018robustness, goodfellow2014explaining}. The presented detailed and systematic study of defence and transferability of quantum and classical ML networks in the presence of adversarial attacks not only provides key new insights, but also reveals a potential for future quantum advantage in ML applications, which could be unlocked in the next few years with the anticipated scale-up of quantum processors coupled with the development of sophisticated error correction schemes. 
\\[4pt]
\noindent
In this work, we carried out a systematic set of quantum and classical simulations across a range of image datasets~\cite{726791, xiao2017fashion, krizhevsky2009learning, yang2015facial} and by creating a variety of adversarial attacks. Our results analyse and compare both the defence of classical(quantum) networks against quantum(classical) adversarial attacks, and the transferability of adversarial examples within classical and quantum ML methods in black box settings. The focus of our work is on comparing the performance of quantum variational classifiers (QVCs), classical convolutional neural networks (CNNs) and the well-known classical neural network architecture ResNet18~\cite{he2016deep}. Our results reveal a surprising discovery that, while the adversarial examples constructed by carrying out white box attacks on the QVCs tend to transfer well to the classical networks, the converse is not true, with the QVCs displaying a remarkable resilience to the classical adversarial attacks (see Figure~\ref{fig:1}) in a black box setting. Based on the analysis of perturbations generated from classical and quantum attacks, we interpret that the observed difference between classical and quantum defence mechanisms arises because the QVCs learn a different, but highly meaningful, set of features to the classical networks, which rely on informative but non-robust features of the data. We also investigate the performance of adversarial training of quantum networks under both white box and black box settings, which provides an important new insight that while highly successful for classical networks \cite{wong2020fast, bai2021recent}, the benefits of adversarial training are quite insignificant in further improving the performance of already resilient quantum ML networks against classical attacks. Interestingly, against the quantum attacks, the computationally expensive adversarial training improves the accuracy of quantum ML networks in a white box setting, but offers diminishing return in black box scenarios. Finally, we propose a novel adversarial attack detection technology which relies on classical and quantum ML frameworks both working together to rapidly identify the presence of an adversarial attack. Overall, our results have established that a future deployment of quantum ML solutions in security conscious practical applications could offer a new kind of advantage in terms of robustness of ML frameworks against adversarial attacks, which will be in addition to commonly sought enhancements in speed and/or accuracy. 
\\ \\
\noindent%
\section{II. Results and Discussion}
The testing and benchmarking of QAML models involve selection of datasets, generation of adversarial attacks, implementation of classical and quantum ML models and a systematic investigation of attack transferability and defence. In this work, we investigate QAML across a diverse set of well-known image datasets, including both grey-scale (MNIST~\cite{726791} and FMNIST~\cite{xiao2017fashion}), and RGB colour (CIFAR~\cite{krizhevsky2009learning} and Celeb-A~\cite{yang2015facial}) images. While we use all ten classes of the MNIST and FMNIST datasets, although not a limitation of our work, we restrict to binary classification (ships vs trucks) in the case of CIFAR in order to reduce the computational burden of training the large quantum classifiers. The classification challenge considered for the Celeb-A dataset is to determine whether the pictured person has black hair. Example images from each of the datasets are shown in Figure~\ref{fig:1} (a). After the selection of datasets, we implemented three different types of adversarial attacks: PGD~\cite{pgd}, FGSM~\cite{fgsm}, and AutoAttack~\cite{auto}. These are some of the strongest attacks commonly used in the classical ML literature to test and benchmark the adversarial vulnerability of classical neural networks. On the quantum side, our focus is on standard quantum variational classifiers (QVCs), while on the classical side we consider convolutional neural networks (henceforth labelled as ConvNet for simplicity) and the well-known neural network architecture ResNet18 (henceforth labelled as ResNet for simplicity)~\cite{he2016deep}. The architectures of ConvNet, ResNet, and the QVCs are schematically shown in Supplementary Figure S1. We load the images into the QVCs with the method of amplitude encoding~\cite{larose2020robust}, which can access the entire,  exponentially large, Hilbert space of the quantum computer. We are therefore able to encode the 28$\times$28 grey-scale images of the MNIST and FMNIST datasets into 10 qubits, and the 3$\times$32$\times$32 RGB images of CIFAR and Celeb-A into 12 qubits (as $28\times 28 < 2^{10}$ and $3\times 32\times 32 < 2^{12}$). After encoding, the resulting quantum state is processed by a variable number of layers of parametrised (trainable) single-qubit rotations and entangling CZ gates (refer to Figure~\ref{fig:1}(c) and Supplementary Figure S1 (c)). The QVC networks are labelled based on the number of layers in the architecture, e.g., QVC200 consists of 200 layers. Both classical and quantum networks were rigorously trained to achieve high accuracies (see details in Supplementary Table S2 and Supplementary Figure S3). We note that the learning accuracies of QVCs trained in our work are quite close to the outcomes from the classical networks even for complex RGB datasets (CIFAR and Celeb-A) and despite that the classical networks utilise significantly more resources than the QVCs. Furthermore, our primary aim in this work is to evaluate the robustness of quantum ML in the presence of adversarial attacks, and not so much on its performance for classification tasks, hence QVC parameters were not fine tuned for this purpose. Further details pertaining to the QVC architectures, CNNs, ResNet, and their training procedures are provided in the Appendix. \\[6pt]
\begin{figure*}[ht]
 \begin{center}
 \includegraphics[width=0.95\textwidth]{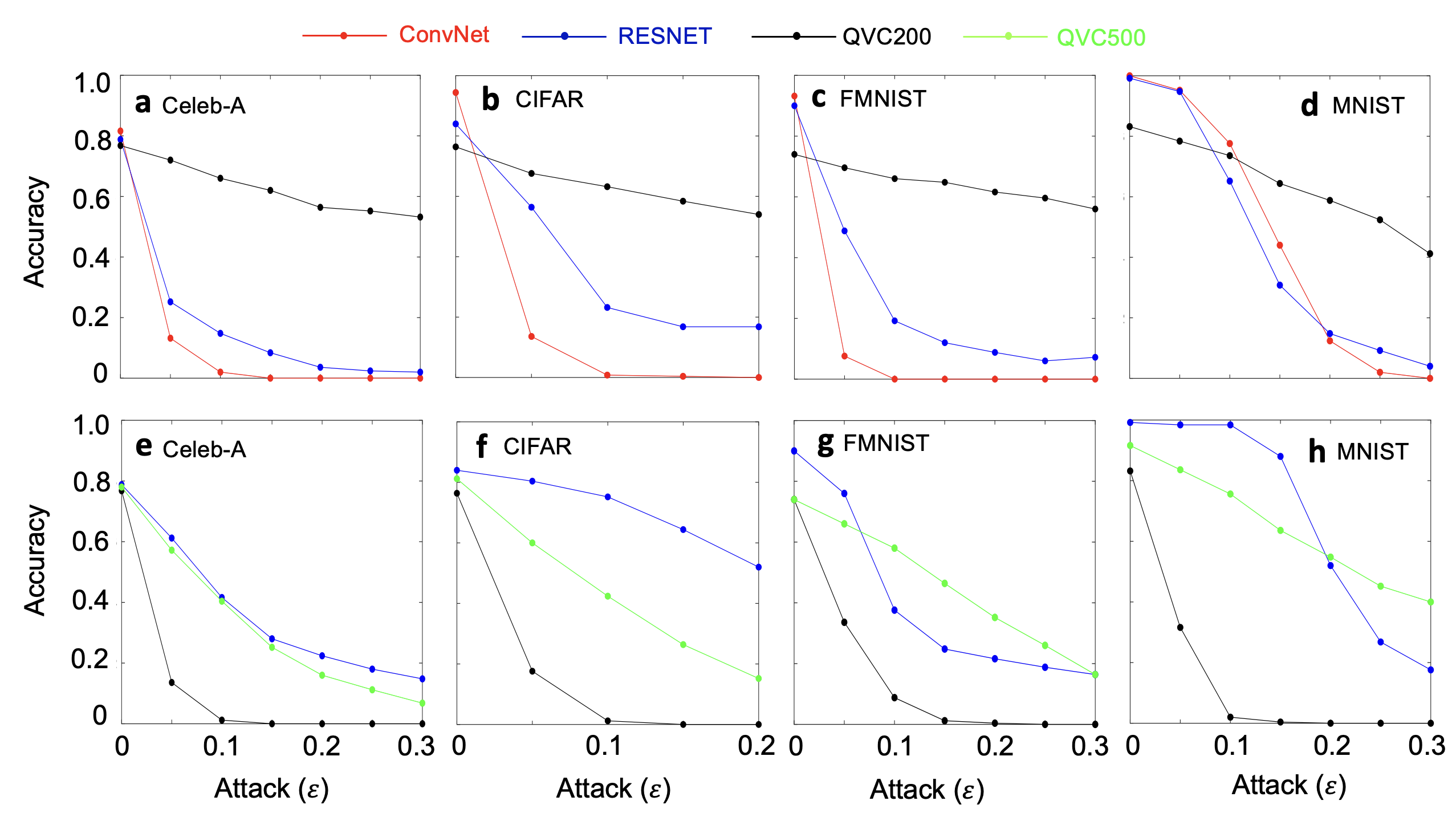}
   \caption{\label{fig:2}\textbf{Transferability and Defence.} The accuracy achieved by classical and quantum networks on sets of 250 adversarially attacked test images from each of the considered datasets in the cases of white box PGD attacks on the convolutional network (top row), and 200 layer quantum variational classifier (QVC200, bottom row) as a function of attack strength $\epsilon$ (measured with the $l_{\infty}$ norm). In both cases we see the accuracy of the network under attack decrease sharply. The tendency of the accuracy of the independent networks to also decrease is a manifestation of the transferability of adversarial examples - they are typically capable of fooling even networks which they were not explicitly designed to attack. We see an exception to this in the top row, with the quantum classifier largely resisting the attacks generated with respect to the convolutional neural network. 
   }
 \end{center}
 \end{figure*}
\noindent
\subsection{A. Adversarial Attack Transferability and Defence} 
A complete testing and benchmarking process of classical and quantum ML networks involves two steps: (1) Adversarial examples created by various white-box attacks on a classical network, e.g., ConvNet and their application to a different classical network, e.g., ResNet, which analyses the transferability of attacks, as the attack generation and testing is done on two entirely different architectures. In the quantum case, this could be performed by generating adversarial examples on, e.g., QVC200 network and assessing its transferability to a QVC500 network. (2) Adversarial examples created by various white box attacks on a classical network such as ConvNet and their application to a quantum network (such as QVC200) which evaluates the defence of a quantum network against that classical attack. This also involved the testing of the defence of the classical networks (ConvNet and ResNet) against the attacks generated from quantum networks such as QVC200. Supplementary Figure S4 provides a simple illustration that defines defence mechanisms and transferability of attacks across classical and quantum networks.   
\\[6pt]
\noindent
Figure~\ref{fig:2} plots our results benchmarking the transferability and defence of both classical and quantum networks for all four datasets in the presence of PGD attacks. The corresponding results for FGSM and AutoAttack are plotted in Supplementary Figures S5 and S6, respectively, which exhibit very similar trends. Let's first discuss the transferability of adversarial attacks (from one classical network to another classical network, and from one quantum network to another quantum network). Along the top row (a-d) of Figure~\ref{fig:2}, we witness the well-documented transferability of adversarial examples between independent classical networks: adversarial examples created by various white box attacks on ConvNet transfer well to ResNet, despite it having an entirely different architecture. Being under white box attack, the accuracy of the ConvNet itself falls quickly (the red lines). The accuracy of an independent classical network, ResNet, also falls quite rapidly (the blue lines), demonstrating the transferability of adversarial attacks in the classical setting, and as seen in Figure~\ref{fig:2} (a-d).
\\[6pt]
\noindent
Along the bottom row (e-h) of Figure~\ref{fig:2} we show, similarly, a successful transfer of adversarial examples generated by a white box attack on QVC200 to an independently trained QVC500 network, although the transferability across quantum networks is relatively weak compared to the classical case. This is perhaps due to the fact that both QVC200 and QVC500 share the same architectural design, with the only difference being the circuit depths, or number of layers. In the future, it would be interesting to test transferability of quantum networks with different architectural designs, e.g., how well the attacks generated from a QVC network transfer to a quantum convolutional network. Nevertheless, Figure~\ref{fig:2} (e-h) show that the accuracy of QVC200 itself falls rapidly in the face of the adversarial examples which are being generated specifically with respect to it (the black lines), as does that of QVC500 (the green lines), demonstrating the successful transferability between two QVC networks. While the transferability of adversarial examples between QVCs has to our knowledge not been studied in the literature, we consider this result unsurprising: QVC200 demonstrates a significant vulnerability to white box adversarial attack, consistent with theoretical expectations~\cite{quantum_cm} and previous empirical studies~\cite{lu2020quantum, ren2022experimental}, and the successful transfer of these attacks to QVC500 demonstrates that the quantum networks are independently utilising similar sets of non-robust features, just as their classical counterparts do~\cite{ilyas2019adversarial}, although to a weaker extent.\\[6pt]
\begin{figure*}[ht]
 \begin{center}
 \includegraphics[width=15.5cm]{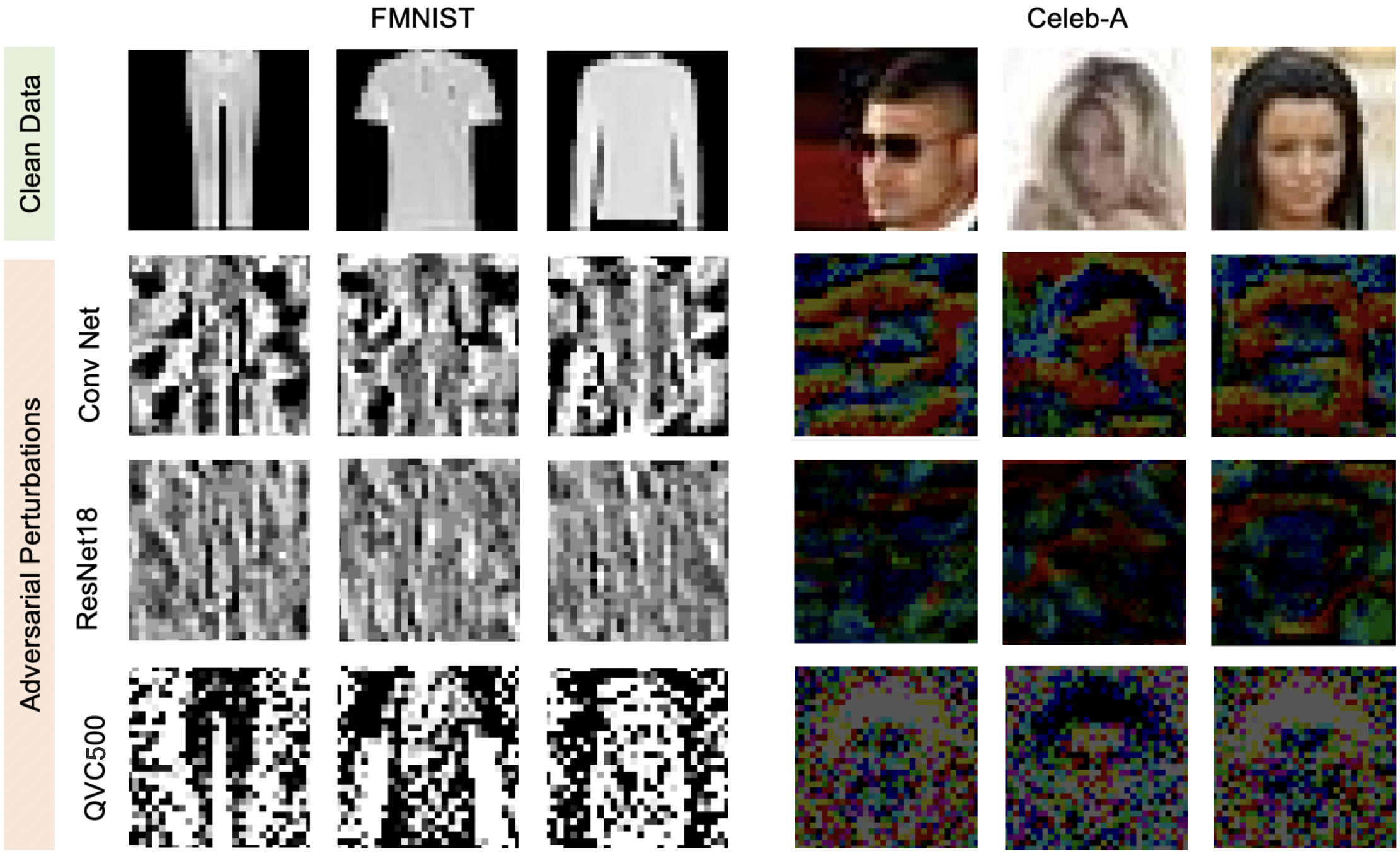}
   \caption{\label{fig:3}\textbf{Content of Adversarial Perturbations.} The adversarial perturbations generated by $\epsilon =0.1$ PGD attacks on the convolutional neural network, ResNet and the 500 layer quantum variational classifier (QVC500) are shown for several examples from the FMNIST and Celeb-A databases. In the case of FMNIST the task is to distinguish between classes of images representing various types of clothing, \textit{e.g.}, t-shirts, pants, shoes, while in the case of Celeb-A the classification problem is to determine whether or not the featured person has black hair. The perturbations generated by the classical networks are incomprehensible to humans, exploiting the highly abstract features discovered by these networks. The attack on the QVC, on the other hand, yields perturbations whose semantic content is clear; they constitute concrete steps towards actually changing the label of the clean image (\textit{i.e.}, filling in the gap between the legs of the pants, adding sleeves to the t-shirt, removing sleeves from the jumper, changing the hair colour of the imaged person). }
 \end{center}
 \end{figure*}
\noindent
Next, we turn our attention to the defence of classical networks against a quantum attack and vice versa. As exhibited by the results plotted in Figure~\ref{fig:2} (a-d), in the presence of PGD attacks generated from the classical ConvNet for various datasets, the QVC200 network demonstrates far superior robustness, retaining reasonable accuracies even in the face of very strong attacks, i.e., $\epsilon \geq$ 0.2. Conversely, the results plotted in Figure~\ref{fig:2} (e-h) show a failure of the classical networks to maintain their accuracy in the face of adversarial examples generated by attacks on the QVC200 networks. Similar conclusions can be drawn from Supplementary Figures S5 and S6, where classical and quantum defences are tested in the presence of adversarial examples generated by FGSM and AutoAttack. One might have expected, a priori, that the classical and quantum networks would learn different features of the data, and that therefore neither attack would transfer well across the classical/quantum divide. The success of the quantum adversarial examples in deceiving the classical networks, however, necessitates  a more careful explanation, which will be discussed in the next section based on an analysis of the underpinning adversarial perturbations in the attacked images in both the classical and quantum realms. 
\\[6pt]
\noindent
\subsection{B. Adversarial Perturbations} 
To gain further insight into the success of the quantum networks in resisting the classical adversarial attacks, and conversely the failure of the classical networks to do the same, we plot the adversarial perturbations generated by PGD attacks on ConvNet, ResNet, and the 500 layer QVC on examples from the FMNIST and Celeb-A datasets in Figure~\ref{fig:3}. Similar examples from the MNIST and CIFAR datasets are shown in Supplementary Figure S7, along with perturbations from the FGSM attack in Supplementary Figures S8 and S9. These perturbations highlight a very similar trend. The perturbations generated in the attacks on the classical networks display complicated high-frequency patterns, which are not readily human understandable. The fact that these ``worst case'' perturbations (from the networks' perspective) take such a form is indicative of the fact that the classical neural networks have learnt to classify the images by identifying extremely complicated patterns in the data rather than following the data distributions, or in other words using a more human-like recognition of the large scale features in the images. While these features may be highly informative, the fact that classical networks tend to independently discover similar sets of them contributes to the vulnerability of the networks to transferred attacks, as the features can then be simultaneously targeted by an adversarial attack on any one of the classifiers individually. The inability of the classical attacks to transfer to the QVCs then indicates that the quantum models are utilising a different set of features. 
\begin{figure*}[ht]
 \begin{center}
 \includegraphics[width=\textwidth]{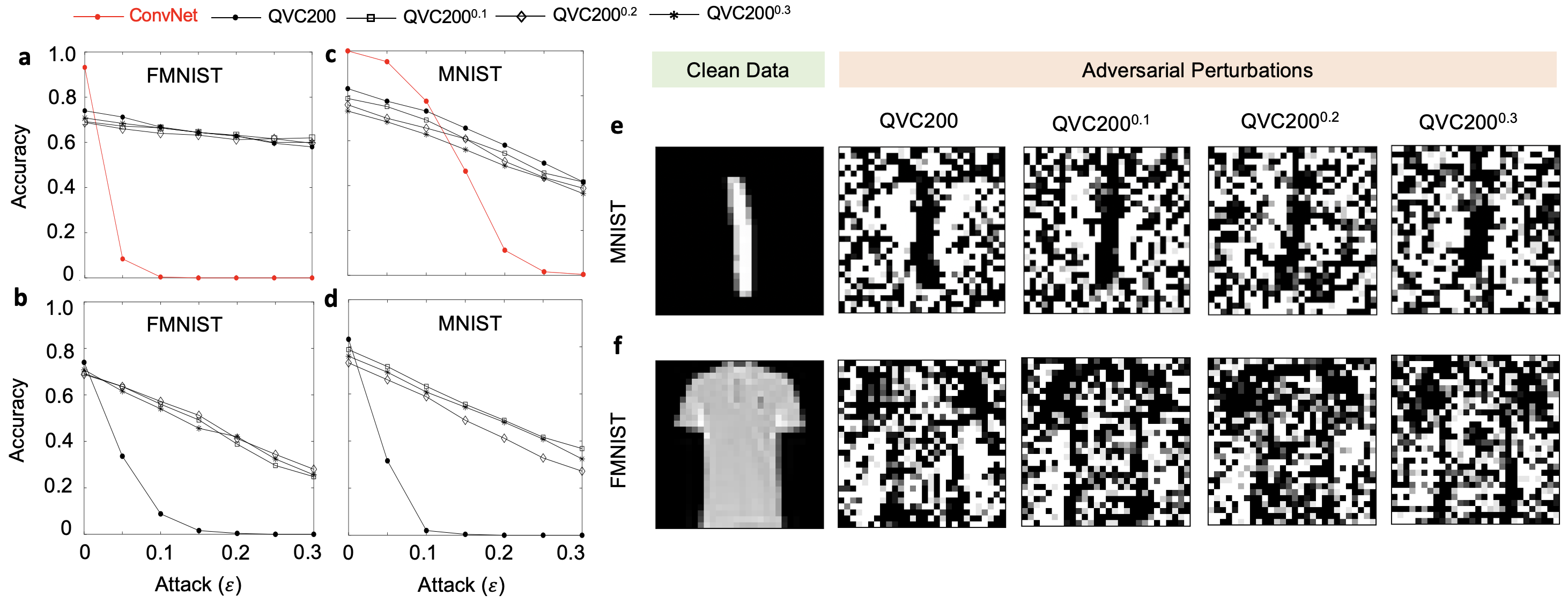}
   \caption{\label{fig:4}\textbf{Quantum ML Adversarial Training.} \textbf{a.} The accuracies achieved by quantum ML networks are plotted as a function of attack strength on a set of 250 adversarially attacked test images from the FMNIST dataset in the cases of white box attacks on the ConvNet. The attack is applied to QVC200 and adversarially trained QVC200 networks (QVC200$^{0.1}$, QVC200$^{0.2}$, and QVC200$^{0.3}$ where adversarial training is performed with PGD attacks of $(l_{\infty})$ strength 0.1, 0.2 and 0.3 respectively). The adversarial training makes a negligible difference to the accuracy of the quantum network against classical attacks. \textbf{b.} The accuracies achieved by quantum ML networks are plotted as a function of attack strength on a set of 250 adversarially attacked test images from the FMNIST dataset in the cases of white box attacks on the QVC200. The attack is applied to QVC200 and adversarially trained QVC200 networks (QVC200$^{0.1}$, QVC200$^{0.2}$, and QVC200$^{0.3}$ where adversarial training is performed with PGD attacks of $(l_{\infty})$ strength 0.1, 0.2 and 0.3 respectively). The adversarial training significantly improves the accuracy of the quantum network.  \textbf{c,d.} The accuracies achieved by ConvNet, ConvNet$^{0.1}$, ConvNet$^{0.2}$ , ConvNet$^{0.3}$, QVC200$^{0.1}$, QVC200$^{0.2}$, and QVC200$^{0.3}$ on the FMNIST dataset in the case of white-box attacks on the QVC200. The classical networks exhibit high accuracy when adversarially trained. As before, adversarial training makes negligible difference to the accuracy of the QVC200. \textbf{e, f.} The adversarial perturbations generated by PGD attacks on QVC200, QVC200$^{0.1}$, QVC200$^{0.2}$, and QVC200$^{0.3}$ are shown for a sample image from the MNIST and FMNIST datasets. The perturbations corresponding to QVC200 show the presence of clear features which actually change the label of the clean image (\textit{i.e.} adding sleeves to the t-shirt). The perturbations from adversarially trained QVCs are very similar to the QVC which underwent standard training.}
 \end{center}
 \end{figure*}
\noindent
Indeed, the perturbations generated by attacking QVC200 demonstrate a markedly different story: in each case we can clearly identify meaningful information in the perturbation. For example, in the first column of Figure~\ref{fig:3}, the perturbation is filling in the gap between the legs of the pants, which would indeed alter the true label of the image were the perturbation strong enough. Similarly, in the second column we see the attack adding sleeves to the T-shirt, and in the third column removing them. On the right hand side, where the classification task is to determine whether or not the pictured person has black hair, we can again see meaningful perturbations generated by the QVC, either lightening or darkening the hair of the pictured person so as to change the label of the data sample image. The fact that, in the case of the QVC, the ``worst case perturbations'' do genuinely correspond to efficiently changing the true label of the sample, rather than incomprehensible noise as is the case for classical networks, implies that the QVC is learning more meaningful patterns in the data, and so is not easily fooled except by meaningful perturbations. This discovery is reminiscent of a similar phenomenon observed in the case of (classical) adversarially trained classifiers, where the adversarial perturbations of classifiers which have been explicitly designed to be robust display more meaningful structure~\cite{ilyas2019adversarial, tsipras2018robustness} than those which have not. We will discuss adversarial training in more detail in the next section. Remarkably, the ability of the QVCs to display similar behaviour despite having undergone only standard (non-adversarial) training is highly interesting and is an indication of a novel kind of possible quantum advantage in ML tasks.
\\[6pt]
\noindent
The qualitative differences between the classical and quantum perturbation landscapes explain both the success of the QVC adversarial examples in transferring to the classical networks, and the failure of the adversarial examples generated by attacking ConvNet to transfer to the QVCs. The classical networks rely on non-robust, but highly informative features which allow them to achieve high accuracy on clean data, but are highly susceptible to adversarial attacks. As the QVCs do not seem to rely on such features to the same extent, an adversarial attack which targets them is limited in its ability to fool the QVCs, but the attacks on QVC200 transfer well to the classical networks by virtue of their meaningful content. 
\\[6pt]
\noindent
These findings suggest extending the distinction drawn in the classical adversarial ML literature~\cite{ilyas2019adversarial, tsipras2018robustness} between robust and non-robust features to include a third category, classically intractable features. While these classically intractable features may themselves be robust or non-robust, we expect them to be robust against classically generated attacks in practice, susceptible at most to perturbations generated by attacking a quantum model. The QVCs studied in this work have discovered features consistent with this classically intractable but non-robust category, being susceptible to attacks transferred from quantum models, but not classical ones. The investigation of quantum models which are also robust against attacks transferred from quantum classifiers is an interesting direction for future work.
\\ \\
\subsection{C. Adversarial Training} 
The field of adversarial ML has seen a long battle between new attacks, proposed defence mechanisms, reformulated counter-attacks, updated defence mechanisms, and so on \cite{goodfellow2014explaining, madry2017towards, athalye2018obfuscated, carlini2017adversarial, miller2020adversarial, guo2017countering, 46641, feinman2017detecting, salman2019provably,zhang2019theoretically}. Throughout this evolving landscape of attacks and defences, an enduring strategy on the defensive side has been that of adversarial training, in which adversarial examples are calculated and included in the training set of the ML model \cite{goodfellow2014explaining, wong2020fast, madry2017towards, bai2021recent}. Despite its simplicity, and the lack of a rigorous guarantee of its success, adversarial training has been found to be effective in practice, and is considered to be one of the strongest methods for building adversarial robustness in classical ML \cite{wong2020fast, bai2021recent}. In this work, we assess the capability of adversarial training for quantum ML networks in the context of the FMNIST and MNIST datasets, with the training-time adversarial examples generated by the PGD, FGSM, and Auto attacks with $\epsilon = 0.1,\ 0.2$ and $0.3$. While a recent study considered adversarial training of a shallow QVC in the context of only binary classification on the MNIST dataset \cite{lu2020quantum}, we note that this is the first time to our knowledge that a deep quantum ML network has undergone adversarial training with complex mutli-class data, \textit{i.e.}, all ten classes of the FMNIST and MNIST datasets in Figure~\ref{fig:4} (a-d). Due to the extensive computational requirements, we leave the adversarial training of the 12-qubit classifiers employed for the CIFAR and Celeb-A datasets to a future study.
\\[6pt] 
\noindent
We plot in Figure~\ref{fig:4} (a-d) the results generated by creating adversarial attacks on ConvNet and QVC200, and applying to adversarially trained QVC200 networks. In (a,c), ConvNet is attacked in a white box setting and the attack is transferred to traditionally trained QVC200 and adversarially trained QVCs (black box scenario). We use superscripts to indicate the strength of the attacks used in the adversarial training,\textit{ i.e.}, the labels QVC200$^x$ indicate adversarial training with perturbations of $l_{\infty}$ norm $x$, where $x \in \{0.1, 0.2, 0.3\}$. Our results show that adversarial training of QVC200 for both FMNIST and MNIST datasets makes only a minor difference in the accuracy of quantum networks. This is in stark contrast to classical ML where adversarial training has been shown to work quite well. Interestingly, it has been reported in the literature that adversarial training of classical ML also significantly reduces their clean learning accuracies (at $\epsilon$=0), but for QVCs no considerable reduction is observed in our work, in particular for the FMNIST data. As the adversarial training is notoriously computationally expensive, our results indicate that in terms of QVC defence against classical attacks, expensive adversarial training is not justified. However, we acknowledge that these insights are based on the conducted simulations and further studies may be needed to establish this conclusion in a fully general set-up. 
\\[6pt]
\noindent
Next, we generate quantum attacks on QVC200 in a white box setting and evaluate the accuracy of adversarially trained QVC200 networks in the presence of those quantum attacks. Figure~\ref{fig:4} (b,d) plots our results for FMNIST and MNIST datasets. In the quantum attack case, we show that the adversarial training improves the accuracy of quantum networks, although it does not completely restore it. We also note that the extent of adversarial training (0.1 or 0.2 or 0.3) only has a little impact, which is again in contrast to the classical ML literature where the level of adversarial training makes a big performance difference. The impact of QVC200 adversarial training in the presence of FGSM and Auto attacks is qualitatively similar to the PGD attacks as indicated from the comparison of results between Figure~\ref{fig:4} and Supplementary Figure S11. Finally, we also tested the resiliency of adversarially trained QVC200s in a black box setting and the results are plotted in Supplementary Figure S12. Here, the attacks are generated in a white box setting on QVC500 and transferred to QVC200 and adversarially trained QVC200s, which indicate that adversarial training only negligibly improves the accuracy. 
\\[6pt]
\noindent
To gain further insights into the adversarial training and its impact on the accuracy of the quantum ML networks, we show the perturbations generated by adversarial attacks on networks both with and without adversarial training in Figure~\ref{fig:4} (e, f). Further examples of such images are shown in Supplementary Figure S13, and Supplementary Figure S11 (e,f) shows examples for FGSM and Auto attacks. The overall conclusions from all these examples is the same that the perturbations generated by attacks on the adversarially trained QVC models display meaningful semantic content and comparable to those generated by attacking the (regularly trained) QVCs. The relatively weak impact of adversarial training on QVC200 accuracy is perhaps due to the fact that, even when undergoing standard training, the QVC networks learnt features quite similar to those they learnt when undergoing adversarial trainings. Also, the perturbation contents from quantum attacks are fundamentally different from the contents of classically generated perturbations (Figure~\ref{fig:3}). Therefore, their susceptibility (or lack thereof) to a given classical attack is similar. This is in contrast to the classical networks, which will default to learning non-robust features unless they are explicitly and rigorously trained in an adversarial setting (see for example Supplementary Figure S10), and therefore adversarial training makes a significant difference.
\\[6pt]
\subsection{D. Prospects for Future Hardware Implementation} 
The QVC results discussed in this work are based on quantum simulations in a noiseless environment. Here we briefly discuss if/when the QVC networks benchmarked in our work can be implemented on quantum processors. The quantum circuits corresponding to the QVC200 networks employed for the MNIST and FMNIST datasets are made up of 200 layers, with each layer consisting of 39 quantum gates (single qubit rotation gates and two-qubit CZ gates), as shown by the schematic diagram in Supplementary Figure S1 (c). The amplitude encoding of the input data into an initial quantum state also requires a few thousand quantum gates. Overall, we estimate that a QVC200 network such as trained in our work would consist of 10$^3$--10$^4$ quantum gates. The current generation of quantum devices are not capable of implementing quantum circuits with such deep circuit depths due to limitations imposed by relatively high level of noise or errors. However, the impact of noise can be analysed to some extent by running quantum simulations with noise models. In Supplementary Figure S14, we have plotted the results obtained from QVC200 on the MNIST dataset in the presence of four different noise models: Depolarisation Noise, Amplitude Damping noise, and Bit-flip errors \cite{note_1}. These noise models are described in the Appendix. Based on the noisy simulations, our results show that the accuracy is independent of noise strength if the noise model is depolarising noise~\cite{du2021quantum} and is only slightly reduced when the bit-flip error rate is increased to 0.5\%. On the other hand, the presence of the Amplitude Damping noise severely impact the accuracy of the developed quantum solution. This is anticipated as the quantum circuit is very deep and therefore the cumulative impact of the noise accumulated over several layers is quite significant. Despite the limitations imposed by noise, the experimental work on the QAML implementation has already begun with a first experimental demonstration just performed in the literature on superconducting quantum hardware \cite{ren2022experimental}, where MNIST and magnetic resonance imaging datasets were used for binary classification using 10-qubits and 26-layers, showing accuracies of above 90\%. Although this study is at the proof-of-concept level, it is already indicating that near-term quantum devices should be able to handle quantum ML problems within the next few years. In our recent survey article, we have discussed the prospects for a surface code \cite{gicev2021scalable} based fault-tolerant implementation of QAML networks and based on our qualitative estimates, we anticipate that a fault-tolerant QAML implementation may be possible on 4000-qubit devices which are expected to become available in 2025 \cite{note_1}. The theoretical analysis presented in our work has established a clear pathway for quantum advantage in QAML and its future fault-tolerant implementation will be a key milestone in the field of quantum computing, unlocking a quantum advantage for a range of real-world ML applications. 
\begin{figure}[ht]
 \begin{center}
 \includegraphics[width=7cm]{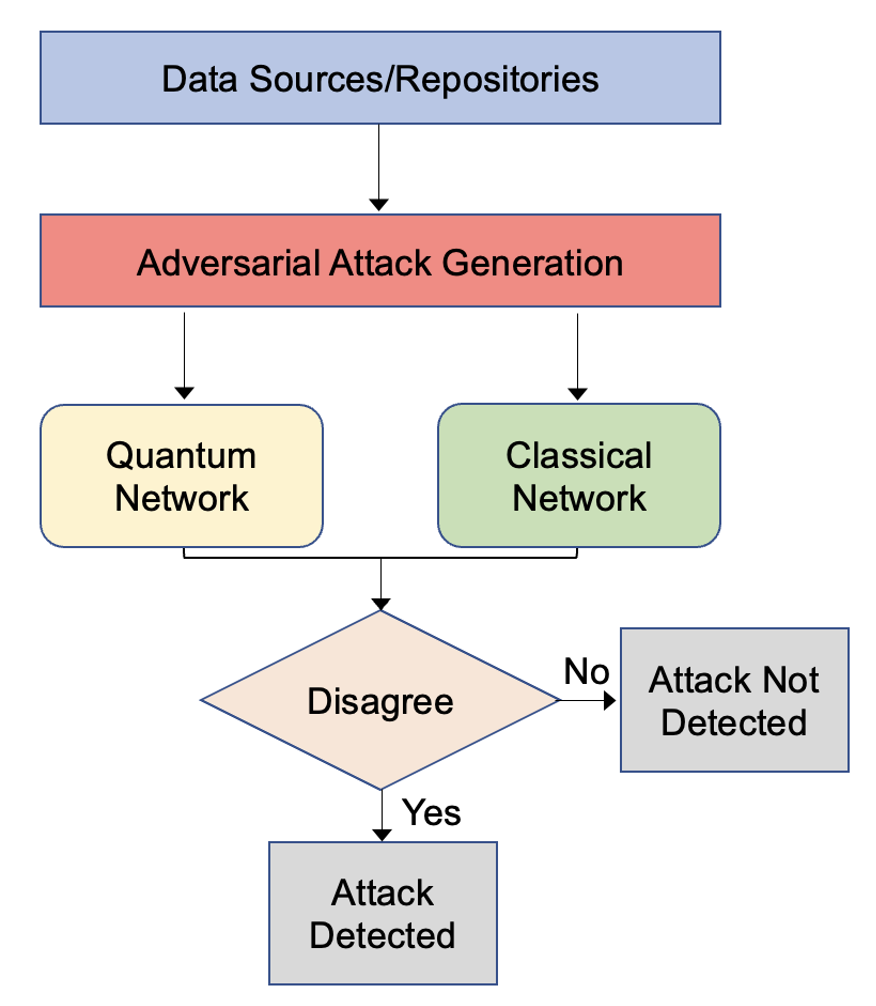}
   \caption{\label{fig:5}\textbf{Attack Detection Technology.} A flowchart diagram illustrating the proposed attack detection strategy based on a hybrid quantum/classical approach. An attack is detected when the predictions of the classical and quantum networks disagree. A future technology developed based on such a combination of classical and quantum networks could provide the critical capability of rapid detection of adversarial attacks on ML models.}
 \end{center}
 \end{figure}
\\[6pt]
\noindent
\subsection{E. Adversarial Attack Detection}
\noindent
The results plotted in Figure~\ref{fig:2}, Supplementary Figures S5 and S6 show that quantum networks defend remarkably well in the presence of classical adversarial attacks. Based on these simulations, we formulate an adversarial attack detection strategy which will be highly useful in practical settings where a rapid detection of adversarial attacks is crucial for reliable ML based solutions. The flowchart diagram in Figure~\ref{fig:5} illustrates our attack detection technology which is based on the hybrid operation of classical and quantum networks working in conjunction to detect the presence of an adversarial attack. In this strategy, an attack is detected if the outcomes from a classical (ConvNet or ResNet) and a quantum network (QVC) disagree. We note that the proposed attack detection technology is only conceptually described here as it is limited by the clean accuracy of the quantum ML network and its practical application would require further development and optimisation of QVC networks to attain clean learning accuracies (at $\epsilon$=0) close to 1.0. We tested the working of the technology by generating 1000 clean and 1000 attacked images (for both MNIST and FMNIST cases) with arbitrary attack strengths and types. For the FMNIST case, the comparison of ConvNet and QVC500 resulted in 227 false positives and 736 true positives, whereas the comparison of ResNet and QVC500 led to 234 false positives and 795 true positives. Here, a false positive means an attack is erroneously detected on a clean image and a true positive means an attack is detected on a perturbed image. Ideally, false and true positives were supposed to be 0 and 1000, respectively, but the actual values are consistent with the relatively low clean accuracy of QVC500. In the future, QVC networks will be further optimised to boost their clean accuracies close to 100\%. \\ \\
\section{III. Summary and Outlook}
Our work has performed a systematic and detailed assessment of the QAML robustness in the presence of various attacks and based on an analysis for a variety of datasets. The presented results have revealed a number of important insights into the working of QAML and bridge a critical knowledge gap by answering key open questions such as: how well will QAML fare against strong classical adversarial attacks, to what extent will attacks transfer across the quantum and classical boundaries, and whether promising classical defence strategies such as adversarial training will work for QML networks. Our work will pave the way for future development of the QAML field and may lead to experimental demonstration of quantum advantage for ML tasks.     
\\[6pt]
\noindent
While our study provides many useful insights which are important to fully understand the working of QAML, there still remains more work to be done for its practical implementation targeting real-world applications. Below, we highlight a few important areas for future development: 

\begin{enumerate} [leftmargin=*, itemsep=0em]

\item We have found that the adversarial training of our QVCs produces relatively small improvements in accuracy, compared to those seen classically in the literature, where it significantly improves the accuracy of classical networks. It will be important to investigate this further, in particular by going beyond QVC architectures and by constructing and benchmarking more complex quantum networks such as quantum convolutional networks. This is of great significance because quantum ML networks have been found to be vulnerable against quantum attacks. If adversarial training could work as well for quantum networks as it works for classical networks, it may allow secure universal robustness of quantum networks against both classical and quantum attacks. Another important line of work would be to further optimise quantum ML networks to improve their clean learning accuracies.  

\item While our work has focused on image data, which has been strongly represented throughout the adversarial ML literature, the extent to which similar results may be found on distinct data types (\textit{e.g.}, audio, text, the environment of reinforcement learning agents, and fully quantum data) also remains an interesting question for future investigations which will enable a wider adaptation of QAML solutions.

\item The construction of deeper QVCs is seen to have limited effect on both the initial accuracy (Supplementary Figure S2) and the robustness to the adversarial attacks (see Supplementary Figure S15). This may be due to the generic form of the QVCs considered here not being particularly well suited to image classification. Analogously to how the original fully connected artificial neural networks were supplanted by convolutional neural networks for image classification and feature detection, these general purpose QVCs may be replaced by specialised quantum generalisations of convolutional neural networks~\cite{qcnn} in order to construct models whose performance on image data exhibits more favourable scaling with model size. 

\item The quantum circuits considered in this work are quite large, consisting of 10 to 12-qubits, 200-1000 layers of single qubit rotation gates and entangling $CZ$ gates, for a total of $\mathcal{O}$(10$^4$) quantum gates. The accurate evaluation of such a circuit is beyond the capabilities of the noisy, non-error corrected quantum computers available today. However, it may be possible using sophisticated error correction methods such as surface-code algorithms to implement the proposed QAML solutions on quantum processors in the near future. We qualitatively estimate that such implementations would require about 5000 or more qubits with error rates requirements below 0.5\% or so. While quantum processors with such configurations may become available in the next few years \cite{ note_1}, it would still require significant development to transpile QAML circuits to error-corrected versions directly implementable on a quantum processor. Further optimisation of the data loading step, which in our current work is very expensive, should ease resource requirements for an implementation on a quantum processor, further pushing forward the practical realisability of QAML approaches.   

\end{enumerate}

\noindent
In summary, vulnerability to adversarial examples has recently emerged as a serious issue confronting classical ML algorithms, raising ongoing concerns about their security and reliability when classifying data from untrusted sources. In the case of quantum ML, too, a similar susceptibility exists, and adversarial examples must be reckoned with if the field is to achieve its expected revolutionary potential. We address an important gap in the QAML literature by thoroughly studying the transferability of adversarial examples between classical and quantum neural networks in the context of common image datasets, discovering a surprising one-way resiliency between quantum and classical networks. The failure of the classical adversarial examples to transfer to the QVCs is characterised as a result of the classical and quantum networks learning different features of the various image datasets, and the quantum ML networks being largely impervious to the specialised attacks which targeted the precise features employed by the classical networks. Although demonstrated on various standard image datasets in this work, the consequence of the differences between the architectures of the classical and quantum classifiers which allow quantum ML to learn from classically intractable and robust features will only become more drastic in future large-scale quantum classifiers. In this case we would suspect even more strongly that classically generated adversarial examples will fail to transfer to these powerful quantum classifiers, which are relying on features which are invisible to the classical networks, and therefore cannot be targeted in an attack. Such a scenario will offer a new form of advantage in QML, orthogonal to the commonly anticipated gains in terms of speed or accuracy -- quantum ML classifiers which may not be necessarily more accurate than their classical counterparts, but exhibit superior robustness to adversarial attacks.

\section*{Acknowledgements}
The authors acknowledge useful discussions with Charles D. Hill, Shu-Lok Tsang and Jia S. Low at the University of Melbourne. MTW acknowledges the support of the Australian Government Research Training Program Scholarship. SME is in part supported by Australian Research Council (ARC) Discovery Early Career Researcher Award (DECRA) DE220100680. The authors acknowledge the support from IBM Quantum Hub at the University of Melbourne and Australian Army Research through Quantum Technology Challenge (QTC22). The computational resources were provided by the National Computing Infrastructure (NCI) and Pawsey Supercomputing Center through National Computational Merit Allocation Scheme (NCMAS). The quantum circuit diagram shown in Figure 1 was generated by using the Quantum User Interface (QUI) developed at the University of Melbourne.
\\[6pt]
\noindent
The authors declare no competing financial or non-financial interests.
\\[6pt]
\noindent
MU conceived the initial idea and supervised the project. MTW developed classical and quantum ML frameworks and performed the simulations, with input from MU. All authors contributed to the analysis of the data. MU and MTW wrote the manuscript with input from all authors.  

\section*{Appendix A: Classifier Details}

\noindent
Here we provide various further details regarding the implementation of the classifiers and adversarial attacks employed in this work.
\\[6pt]
\noindent
\textbf{Classical Network Implementations:} Our convolutional neural network begins with three layers of 3$\times$3 filters, containing 64, 128 and 256 feature maps respectively, with 2$\times$2 maxpooling and the ReLu activation function. These convolutional layers are followed by two fully
connected layers, which also utilise the ReLu activation function. The architecture of ResNet18 is as described in \cite{he2016deep}. The networks were implemented in Pytorch~\cite{pytorch} and trained with the Adam optimiser and the cross-entropy loss function \cite{zhang2018generalized}. Further details of the network architectures are shown in the Supplementary Materials.
\\[6pt]
\noindent
\textbf{Quantum Network Implementations:} 
Our QVCs follow a standard three step process for processing the input image data before outputting a predicted label. In the first step the images are encoded into a quantum state via the method of amplitude encoding~\cite{larose2020robust}. In the second step the encoded images are processed by passing through a parameterised quantum circuit consisting of a variable number of layers, with each layer consisting of a parameterised rotation to each qubit followed by nearest neighbour CZ gates. In this work we employ deep circuits with 200, 500 or 1000 layers. Finally, in the third step we measure the $z$ expectation value of the first $m$ qubits, where $m$ is the number of classes in the particular classification problem being considered. The prediction of the QVC is defined to be the index of the highest of these values. Further details may be found in the Supplementary Materials.  The QVCs were implemented in Pennylane~\cite{pennylane} and trained with the Adam optimiser and the cross-entropy loss function \cite{zhang2018generalized}. 
\\[6pt]
\noindent
\textbf{Adversarial Training:} We employ adversarial training with the PGD attack with 3 gradient descent steps. Each batch of training examples consists of 50\% clean and 50\% adversarial examples.
\\[6pt]
\noindent
\textbf{Quantum Noise:} We test our QVCs in the presence of depolarisation noise, amplitude damping, and bit flip noise. The results are shown in the Supplementary Figure S14. Depolarisation Noise: In this model, the state of the system is replaced with the maximally mixed state with probability p which indicates the strength of the noise. Amplitude Damping: This model is a quantum channel which gives a simple model for relaxation of an excited state to the ground state. In the qubit case, with some probability there is a spontaneous transition $\left.|0\right\rangle\rightarrow\left.|1\right\rangle$. The lifetime is governed by a parameter $\gamma$, which determines the strength of the noise. Bit-flip: In this model, an X gate is applied with probability p, which determines the strength of the noise.\\

\def\bibsection{\subsection*{\refname}}
\bibliographystyle{naturemag}
%\bibliography{refs}

\clearpage
\newpage

\setcounter{figure}{0}
\onecolumngrid

\begin{center}
\Large{\textbf{Supplementary Information for \\ Benchmarking Adversarially Robust Quantum Machine Learning at Scale}}
\end{center}

\noindent%
\\ \\
\large{{\textbf{S1. Network Architectures}}}
\normalsize
\\ \\
Our convolutional neural network begins with three layers of 3$\times$3 filters, containing 64, 128 and 256 feature maps respectively,
with 2$\times$2 maxpooling and the ReLu activation function.
These convolutional layers are
followed by two fully connected layers, which also utilise the ReLu activation function. The architecture of ResNet18 is as described in Ref. \cite{he2016deep} and is shown schematically in Supplementary Figure \ref{fig:arch}(a). 
Both the classical and quantum networks are trained with the Adam optimiser and the cross-entropy loss function \cite{zhang2018generalized}. Further details of the network architectures are shown in the table below. 

\begin{center}
\begin{table}[h!]
\begin{tabular}{|c||c|c|c||c|c|c|}
  \cline{2-7}
    \multicolumn{1}{c|}{} & \multicolumn{3}{c|}{MNIST and FMNIST}  & \multicolumn{3}{c|}{CIFAR-10 and Celeb-A}  \\ \cline{1-7}
      Network  &   Qubits &   Parameters &  Layers & Qubits &   Parameters &  Layers \\ \hhline{|=#=|=|=#=|=|=}
      ConvNet  &  -  & $\sim 10^6$ & 5 &  -  & $\sim 10^6$  & 5 \\ \hline
      ResNet18 &  -  & $\sim 10^7$  & 10  &  -  & $\sim 10^7$ & 18 \\ \hline
      QVC200   &  10 & 6000  & 200  &  12 & 7200  & 200 \\ \hline
      QVC500   &  10 & 15000 & 500  &  12 & 18000  & 500  \\ \hline
      QVC1000  &  10 & 30000 & 1000 &  - & - & - \\ \hline
  \end{tabular}
  \caption{The resource requirements for the different networks considered.
In all cases the quantum variational classifiers require drastically less
trainable parameters than their classical counterparts.
Due to computational restrictions QVC1000 was not run on the 12-qubit datasets.}
  \label{table:qvc500}
\end{table}
\end{center}

\large{{\textbf{S2. Introduction to Adversarial Machine Learning}}}
\normalsize
\\ \\
\noindent
We begin with a brief review of the aspects of
adversarial machine learning relevant to our work,
drawing on both the classical~\cite{szegedy2013intriguing} and quantum~\cite{lu2020quantum} supervised machine learning literature.
The key discovery of adversarial machine learning is that standard ML frameworks are highly susceptible to
being deceived by subtle, malicious tampering with their input data~\cite{biggio2013evasion,szegedy2013intriguing,10.1145/2046684.2046692,kurakin2016adversarial}.
The results of such tampering, adversarial examples, can be readily produced
by taking a data sample (which for a high performing ML method will likely be correctly classified)
and attempting to find a tiny perturbation
which when added to the data causes a misclassification.
By insisting that the perturbation is small one guarantees that the true label of the constructed adversarial example is the same
as the label of the original datapoint, despite the model classifying them differently if the attack is successful.
\\ \\
\noindent
Concretely, suppose that we are attempting to train a classifier to label data from a set $\mathcal{X}$, with corresponding labels
from another set $\mathcal{Y}$. The goal of supervised machine learning is then to learn a parameterised function
$ C_{\boldsymbol{\theta}}: \mathcal{X}\to \mathcal{Y}$ in order to minimise the empirical loss obtained on a given training set
consisting of pairs $ (\boldsymbol{x}_i , y_i) $ of labelled examples:
\begin{equation}
 \boldsymbol{\theta} = \argmin_{\boldsymbol{\theta}'\in\Theta} \frac{1}{N} \sum_i \mathcal{L}\left( C_{\boldsymbol{\theta}'}(\boldsymbol{x}_i ), y_i \right)
  \label{eq:loss}
\end{equation}
where $\Theta$ is the set of possible parameter values, and $\mathcal{L}$ is a chosen loss 
function (\textit{e.g.} the cross-entropy loss~\cite{zhang2018generalized}).
Alternately, in the setting of adversarial machine learning, one is
given a trained classifier $C_{\boldsymbol{\theta}}$ and an input sample $\boldsymbol{x}$, and seeks to
construct an adversarial perturbation $\boldsymbol{\delta}_{\mathrm{adv}}$ by maximising the loss function:
\begin{equation}
 \boldsymbol{\delta}_{\mathrm{adv}} = \argmax_{\boldsymbol{\delta}\in\Delta} \mathcal{L}\left( C_{\boldsymbol{\theta}}(\boldsymbol{x} + \boldsymbol{\delta}), y \right)
  \label{eq:adv_attack}
\end{equation}

\noindent
where $\Delta$ is the set of allowable (\textit{i.e.} acceptably small) perturbations.
While the optimisation problem of Equation~\ref{eq:adv_attack} may be highly nonconvex and analytically intractable,
many strong strategies (``types of attacks'') have emerged for tackling it in practice.
In this work we consider three standard attacks from the classical ML literature, namely
projected gradient descent (PGD)~\cite{pgd}, the fast gradient sign method (FGSM)~\cite{fgsm}, and Auto Attack~\cite{auto}.\\
\\ \\
\noindent
An important classification of adversarial attacks is into white box and black box attacks.
In a white box attack, the adversary is assumed to have full access to the network under attack, including
familiarity with its architecture and the values of its weights and biases. As a result, the adversary possesses the ability to
differentiate with respect to an input to the network, and therefore carry out optimisation based attacks such as the
PGD, FGSM and Auto attacks considered here. While such attacks generally have a devastating effect on the accuracy of a network,
the assumption of such intimate familiarity
with the target network may not always be satisfied. A more realistic case is that where the adversary has access to the
target network only through submitting queries, and is therefore forced to attempt a black-box attack.
The feasibility of black box attacks stems largely from the discovery of the remarkable transferability of adversarial examples
between various ML frameworks: adversarial examples constructed to attack a specific target tend to transfer to entirely
independent networks, deceiving them as well~\cite{szegedy2013intriguing, ilyas2019adversarial, tsipras2018robustness, goodfellow2014explaining}. 
This discovery alleviates the need for direct access to a network in order to reliably attack it;
the adversary may construct a network of their own, perform a white box attack on it, and then simply submit the generated
adversarial examples to the target network.
The transferability of adversarial examples between classical and quantum networks is the main topic of study in this work.\\

\noindent%
\large{{\textbf{S3. Introduction to Quantum Machine Learning}}}
\normalsize
\\ \\
The quantum machine learning (QML) models which we will consider throughout this work belong to the category
of quantum variational classifiers (QVCs).
As in the classical case, these models will be parameterised functions $\mathcal{A}_{\boldsymbol{\theta}}: \mathcal{X}\mapsto \mathcal{Y}$,
where we consider an input set $\mathcal{X}$ with associated labels from a further set $\mathcal{Y}$ and
denote a QVC parameterised by parameters
$\boldsymbol{\theta}$ as $\mathcal{A}_{\boldsymbol{\theta}}$.
Our QVCs follow a standard three step process for processing the input image data
before outputting a predicted label. In the first step the images are encoded into
a quantum state. Due to the strong limitations on the numbers of qubits currently available to quantum computers and simulators,
and our need to encode high dimensional image data, we employ the method of amplitude encoding~\cite{larose2020robust}. Having
represented the image by a vector $\boldsymbol{x}$ containing
its pixel values, amplitude encoding is the mapping
\begin{equation}
  \boldsymbol{x} \mapsto \sum_{i=0}^{2^n} x_i \ket{i}
\end{equation}
where the set $ {\{\ket{i}\}}_{i=0}^{2^n-1}  $ forms the computational basis of the Hilbert space. 
As amplitude encoding makes use of
the entire Hilbert space, which has a dimension exponentially large in the number of qubits, 
it can encode a vector $\boldsymbol{x}\in \mathbb{R}^m $
into $ \lceil \log_2 (m) \rceil $ qubits.
This extreme compression allows us to encode images into a manageable number of qubits;
in the case of the MNIST and FMNIST datasets, which consist of 28$\times$28 greyscale
images, only 10 qubits are needed, while for the 3$\times$32$\times$32 RGB images from the CIFAR-10 and CelebA datasets, 12 qubits are
required. In both cases, zeros are appended to the input vectors in order to make their length a power of two.
Having loaded the images into the quantum model in the first stage, in the second stage 
they are processed by passing through a
parameterised quantum circuit, the parameters of which are updated during training to 
minimise the average cross entropy loss
on the training dataset as in Equation~\ref{eq:loss}. We choose parameterised circuits consisting of a variable number of
layers, with each layer consisting of a parameterised rotation to each qubit followed by nearest neighbour CZ gates.
In this work we employ deep circuits with 200, 500 or 1000 layers. 
Finally, in the third stage a set of measurements are performed in order to determine the predicted output label.
In our case we measure the $z$ expectation value of the first $m$ qubits, where $m$ is the number of classes in the particular
classification problem being considered (10 for MNIST and FMNIST, 2 for CIFAR-10 and CelebA).
Given an input $\boldsymbol{x}$, then, the final output of the quantum classifier is
\begin{equation}
  \mathcal{A}_{\theta} (\boldsymbol{x}) = \argmax_{k\leq m} \Tr \Big[ U_{\theta}(\boldsymbol{x}) \sigma_k^z \Big]
\end{equation}
where we denote the unitary representing the action of the parameterised quantum circuit on the input $\boldsymbol{x}$
by $U_{\theta}(\boldsymbol{x}) $.
The procedure is depicted in Figure 1 of the main text.
\\[6pt]
\noindent
The accuracies obtained by the various QVCs on the considered datasets are shown in Supplementary
Figure~\ref{fig:trainaccs}. \\

%\def\bibsection{\subsection*{\refname}}

%\bibliographystyle{naturemag}
%\bibliography{SiDonor}
%\def\bibsection{\subsection*{\refname}}

%\bibliographystyle{naturemag}
%\bibliography{refs}

\clearpage
\newpage

\renewcommand{\thefigure}{\textbf{S\arabic{figure}}}
\renewcommand{\figurename}{\textbf{Supplementary Figure}}

\renewcommand{\thetable}{\textbf{T\arabic{table}}}
\renewcommand{\tablename}{\textbf{Supplementary Table.}}

\begin{figure*}[ht]
 \begin{center}
 \includegraphics[width=\textwidth]{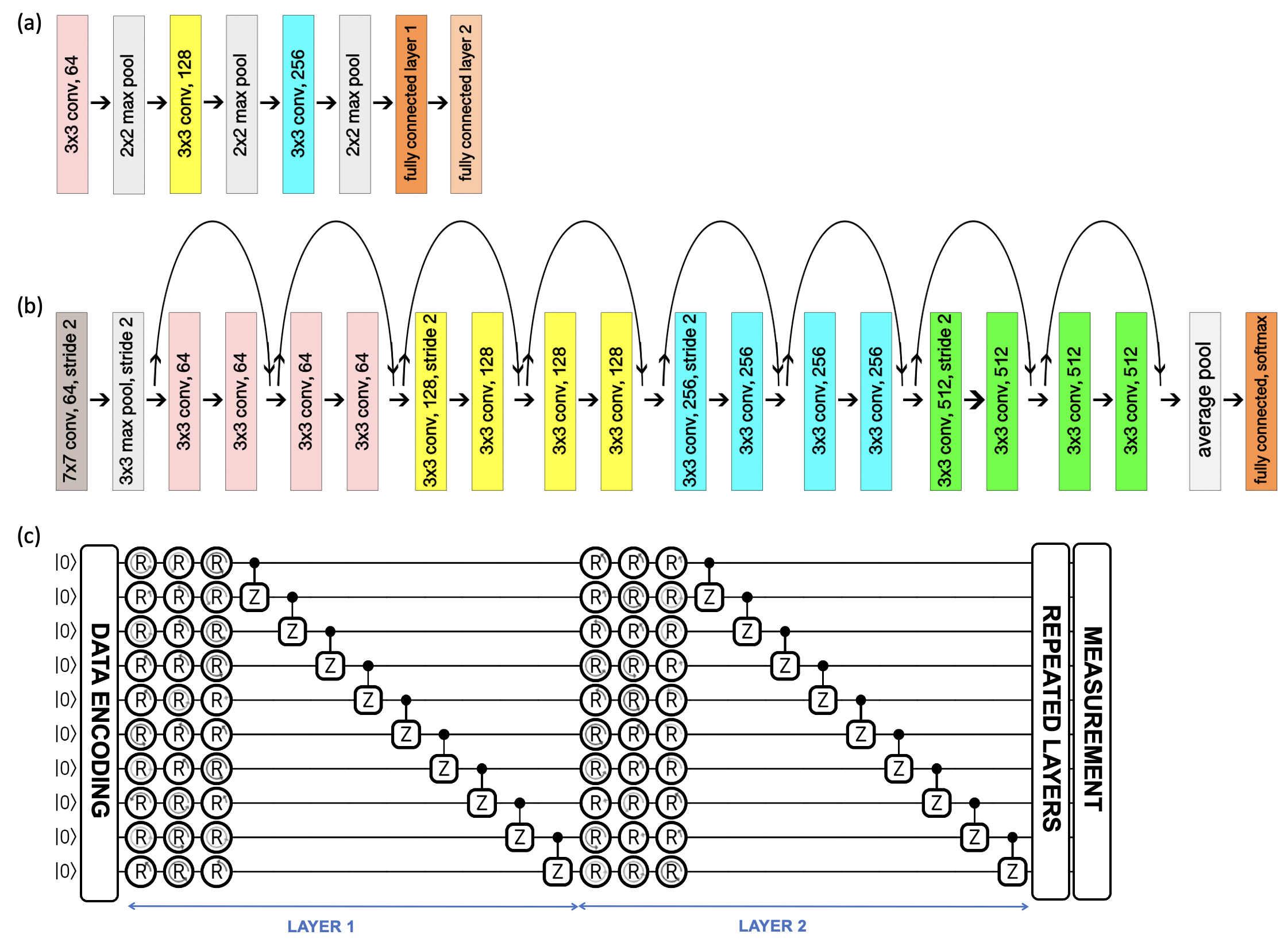}
   \caption{\label{fig:arch} Schematic depictions of the considered architectures are shown for (a) our simple convolutional neural network, (b) ResNet18 (as in  Ref.\cite{he2016deep}), (c) our QVCs. The QVC architecture is discussed in detail in Section S3.  }
\end{center}
\end{figure*}

\begin{figure*}[ht]
 \begin{center}
 \includegraphics[width=0.8\textwidth]{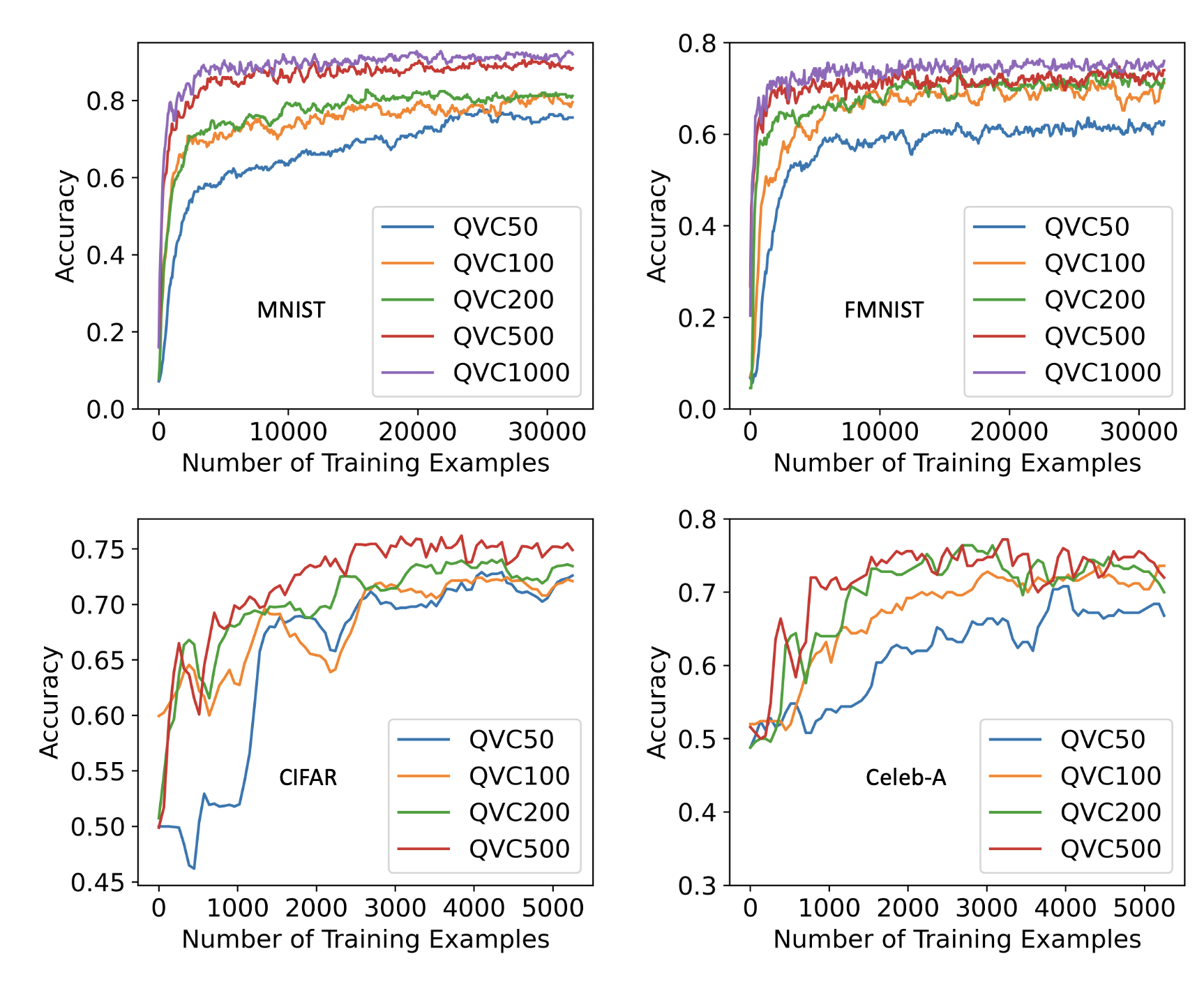}   \caption{\label{fig:trainaccs}The accuracy achieved by the quantum 
   variational classifiers on test sets of 250 images from each dataset throughout the training process. 
   We see consistent but modest gains in accuracy as a function of model size in the cases of MNIST and FMNIST
   (10 class classification problems), but less consistency in the cases of CIFAR-2 and Celeb-A
   (binary classification problems). Architectures more suited to image classification than QVCs (e.g. quantum convolutional 
   neural networks~\cite{qcnn}) may display better scaling behaviour. }
\end{center}
\end{figure*}

\begin{figure*}[ht]
 \begin{center}
 \includegraphics[width=0.8\textwidth]{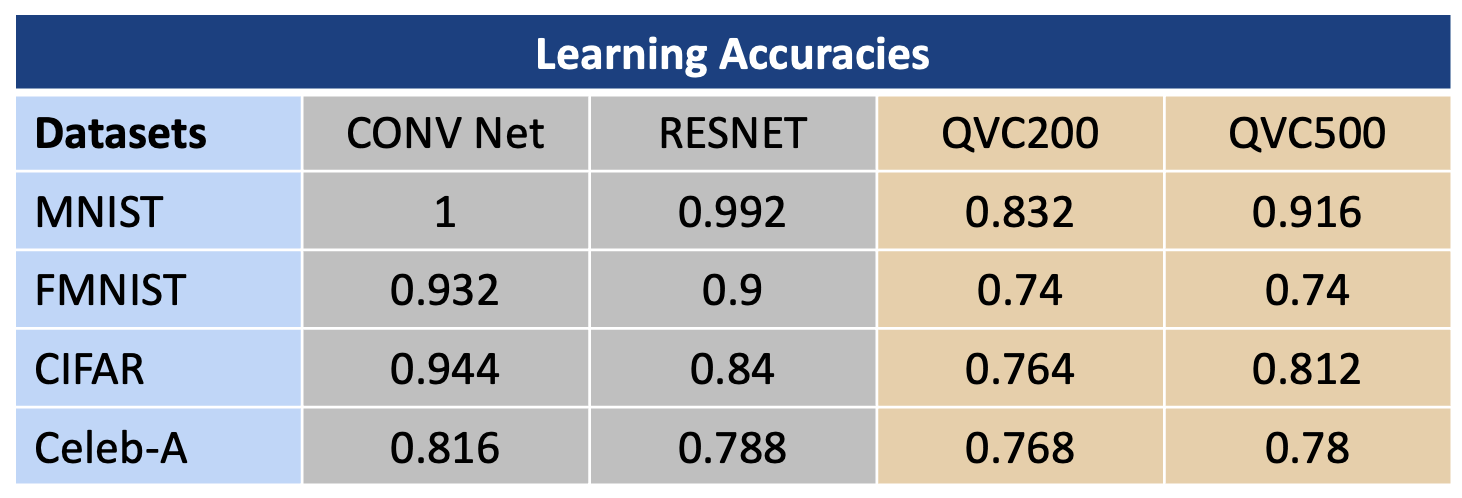}
   \caption{\label{fig:acc} Training accuracies achieved by classical and quantum networks for various datasets.}
\end{center}
\end{figure*}

\begin{figure*}[ht]
 \begin{center}
 \includegraphics[width=0.5\textwidth]{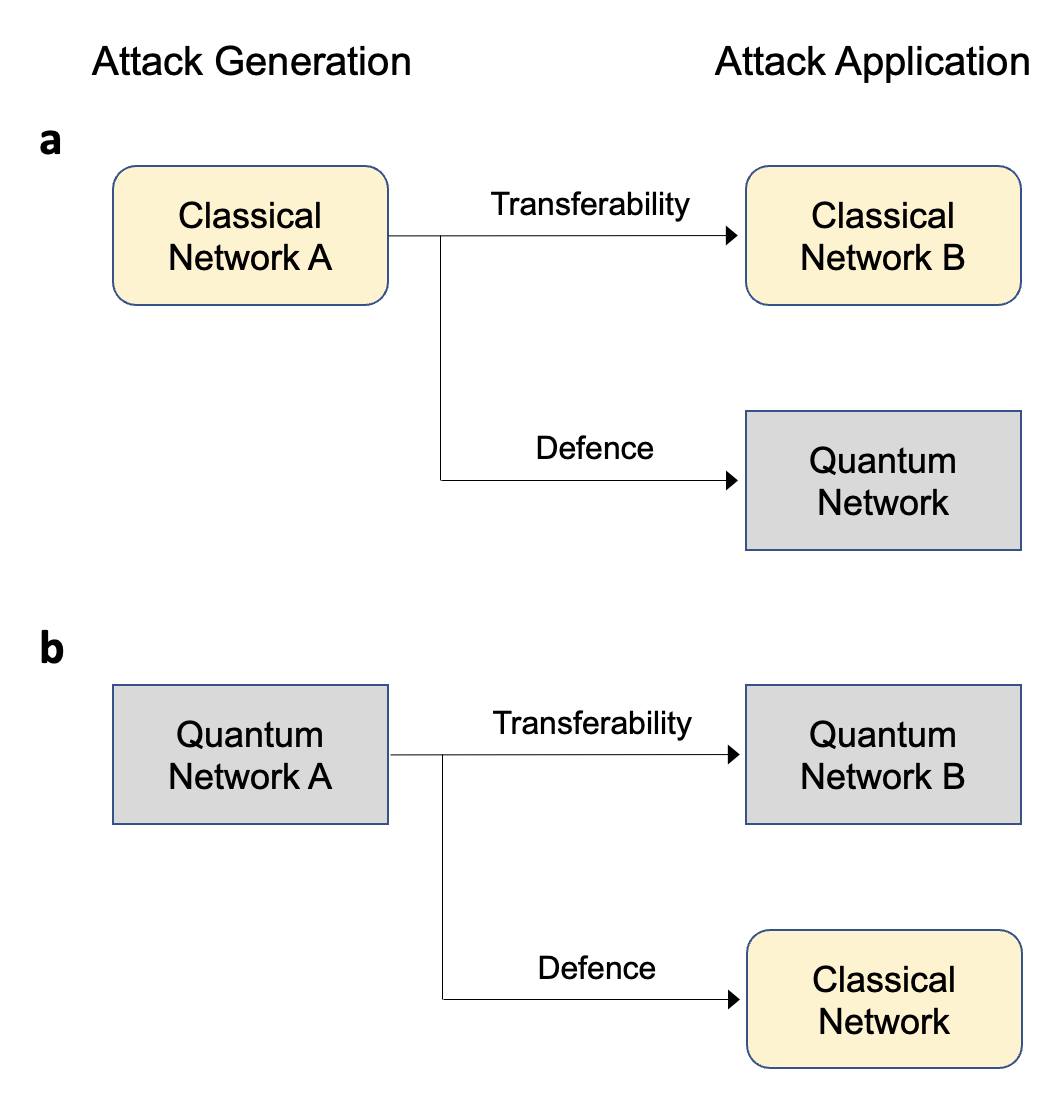}
   \caption{\label{fig:flowchart} A simple illustration to show our definition of transferability and defence. The attack is generated by either a classical network or a quantum network (Attack Generation). The effectiveness of the attack generated by a classical(quantum) network when applied to a different classical(quantum) network is defined as Transferability. The robustness of a quantum(classical) network against a classical(quantum) attack is defined as Defence.}
\end{center}
\end{figure*}

\begin{figure*}[ht]
 \begin{center}
 \includegraphics[width=\textwidth]{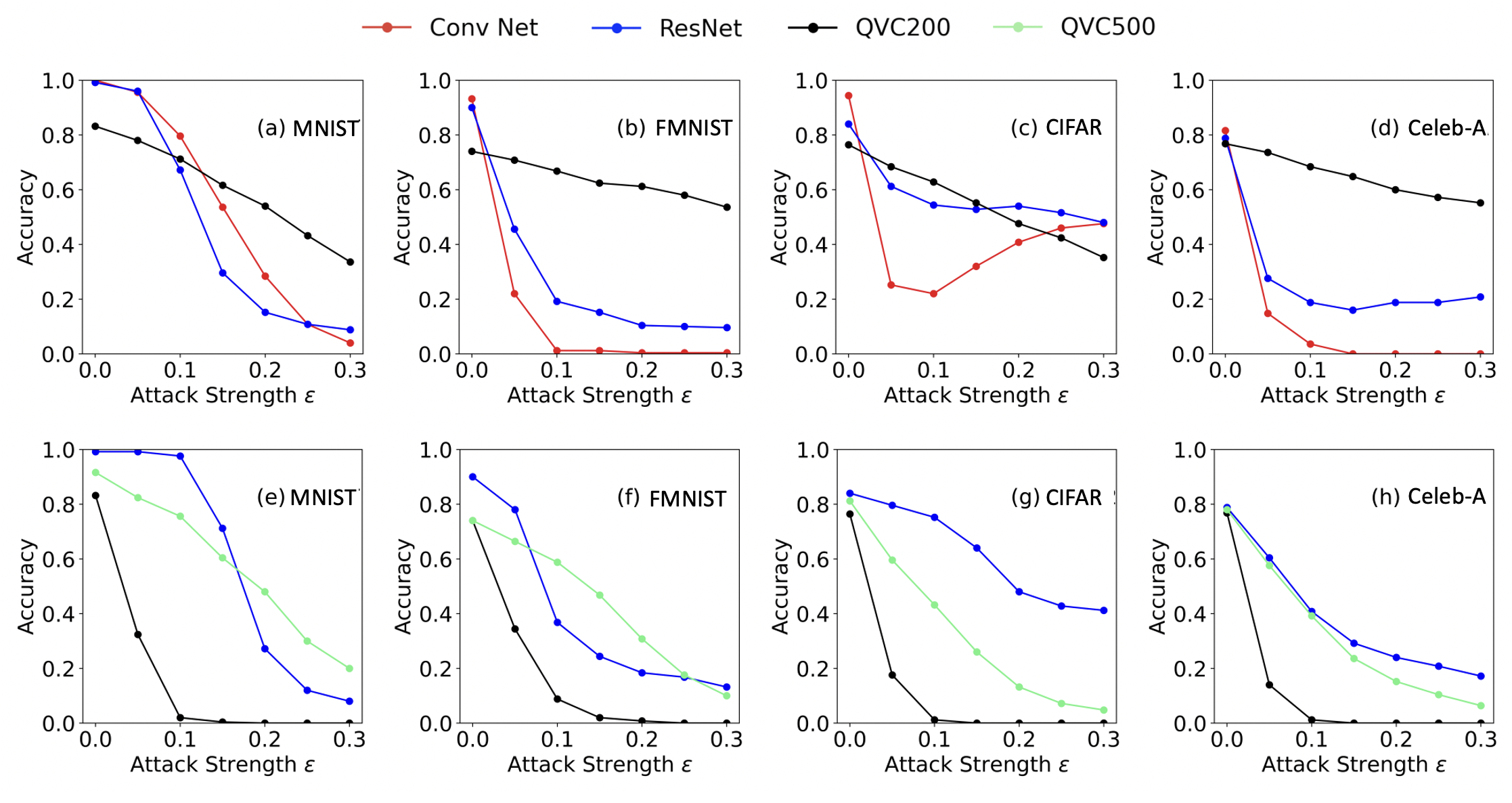}
   \caption{\label{fig:S5} The accuracy achieved by classical and quantum networks
   on sets of 250 adversarially attacked test images from each of the considered datasets
   in the cases of white-box FGSM attacks on the convolutional network (top row), and 200
   layer quantum variational classifier (QVC200, bottom row) as a function of attack strength (measured with the $l_{\infty}$ norm). 
   In both cases we see the accuracy of the network under attack decrease sharply.
   The tendency of the accuracy of the independent networks to also decrease is a manifestation of the transferability
   of adversarial examples - they are typically capable of fooling even networks which they were not explicitly designed to attack. We see an exception to this in the top row, with the quantum classifier usually resisting the attacks generated with respect to the convolutional neural network. 
   }
 \end{center}
 \end{figure*}

\begin{figure*}[ht]
 \begin{center}
 \includegraphics[width=\textwidth]{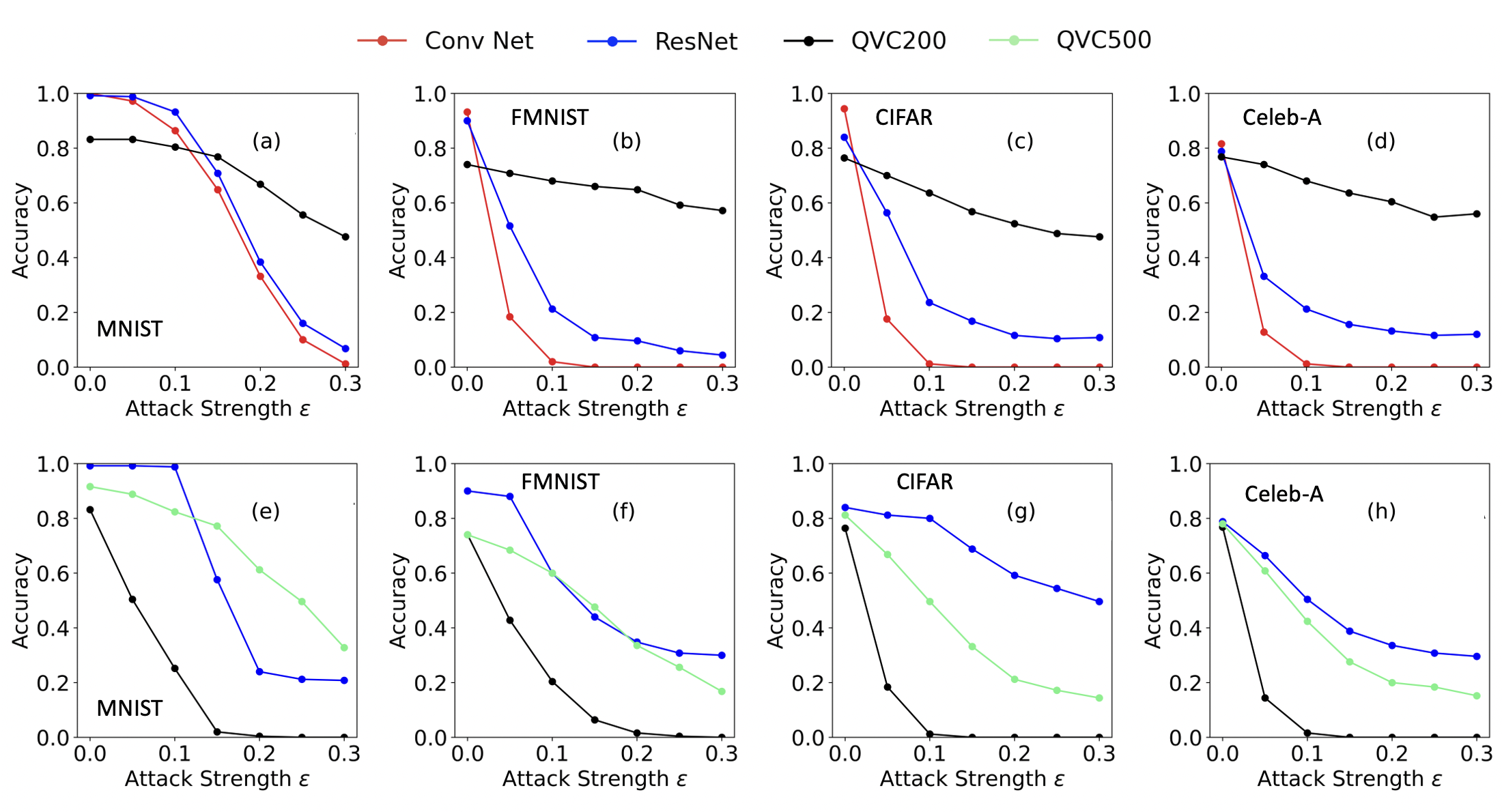}
   \caption{\label{fig:S6} The accuracy achieved by classical and quantum networks
   on sets of 250 adversarially attacked test images from each of the considered datasets
   in the cases of white-box AutoAttacks on the convolutional network (top row), and 200
   layer quantum variational classifier (QVC200, bottom row) as a function of attack strength (measured with the $l_{\infty}$ norm). 
   In both cases we see the accuracy of the network under attack decrease sharply.
   The tendency of the accuracy of the independent networks to also decrease is a manifestation of the transferability
   of adversarial examples - they are typically capable of fooling even networks which they were not explicitly designed to attack. We see an exception to this in the top row, with the quantum classifier largely resisting the attacks generated with respect to the convolutional neural network. 
   }
 \end{center}
 \end{figure*}

\begin{figure*}[ht]
 \begin{center}
 \includegraphics[width=0.8\textwidth]{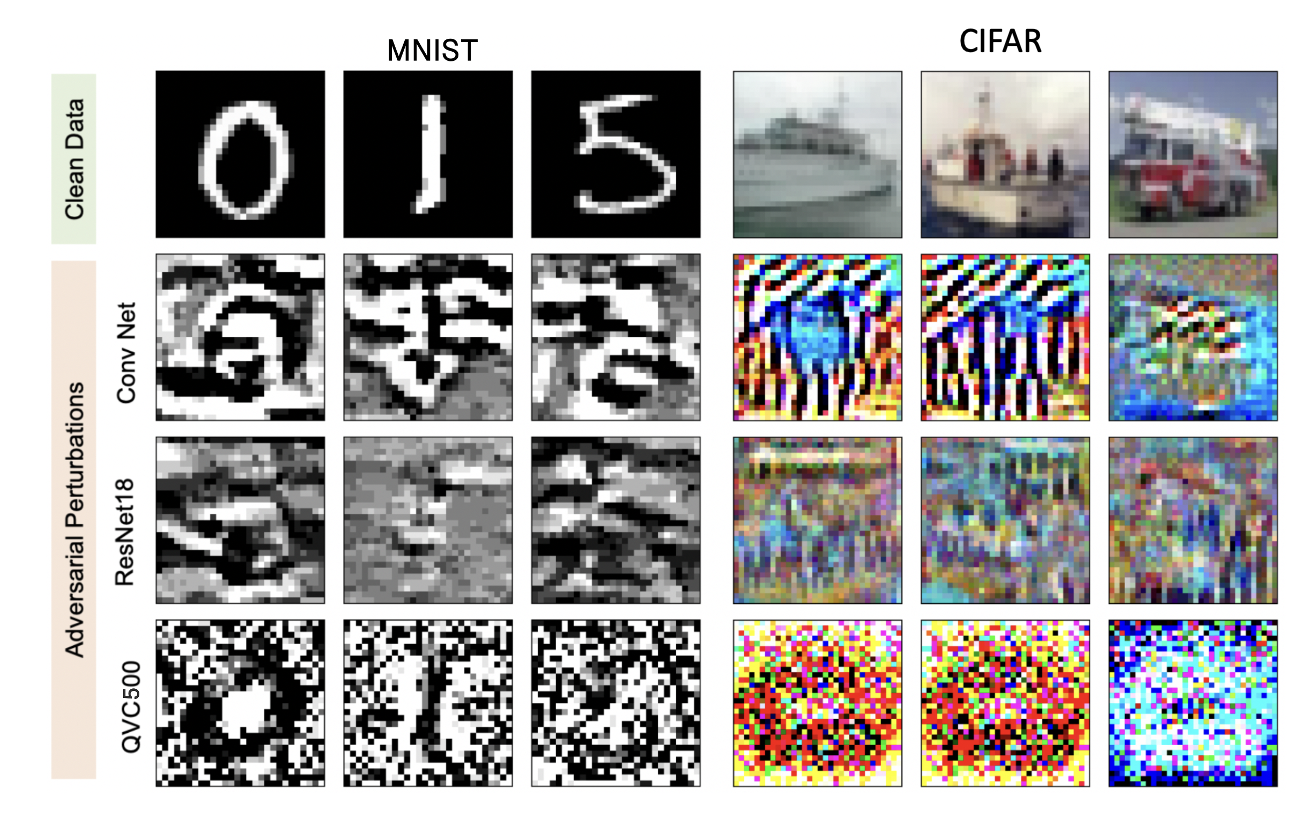}
\label{fig:S7}
   \caption{\label{fig:S6} The adversarial perturbations generated by $\epsilon =0.1$ PGD attacks on
   the convolutional neural network, ResNet18 and the 500 layer quantum variational classifier (QVC500) are shown for several
   examples from the MNIST and CIFAR databases.
   }
\end{center}
\end{figure*}

\begin{figure*}[ht]
 \begin{center}
 \includegraphics[width=0.8\textwidth]{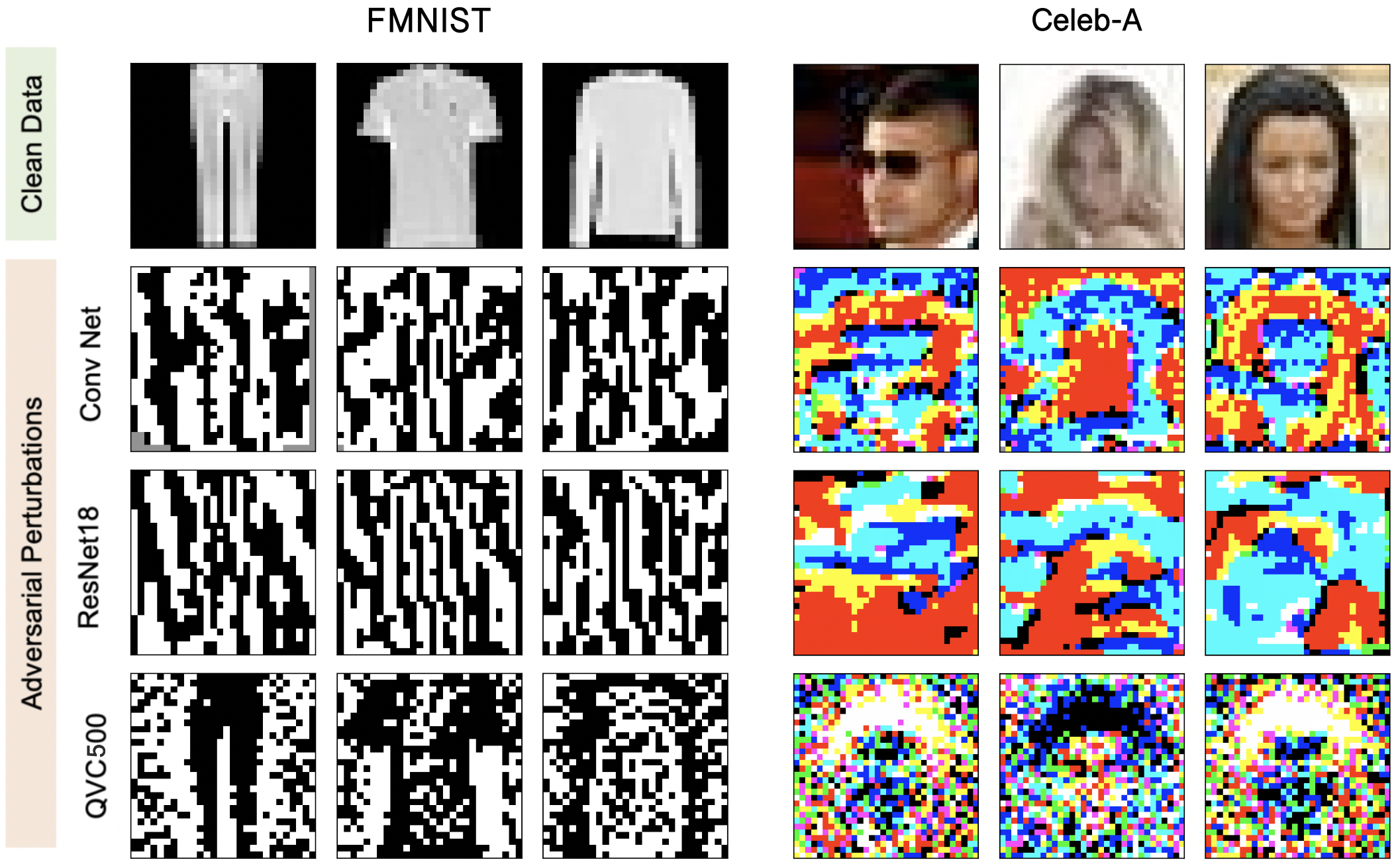}
\label{fig:S8}
   \caption{\label{fig:S8} The adversarial perturbations generated by $\epsilon =0.1$ FGSM attacks on
   the convolutional neural network, ResNet18 and the 500 layer quantum variational classifier (QVC500) are shown for several
   examples from the FMNIST and Celeb-A databases.
   }
\end{center}
\end{figure*}

\begin{figure*}[ht]
 \begin{center}
 \includegraphics[width=0.8\textwidth]{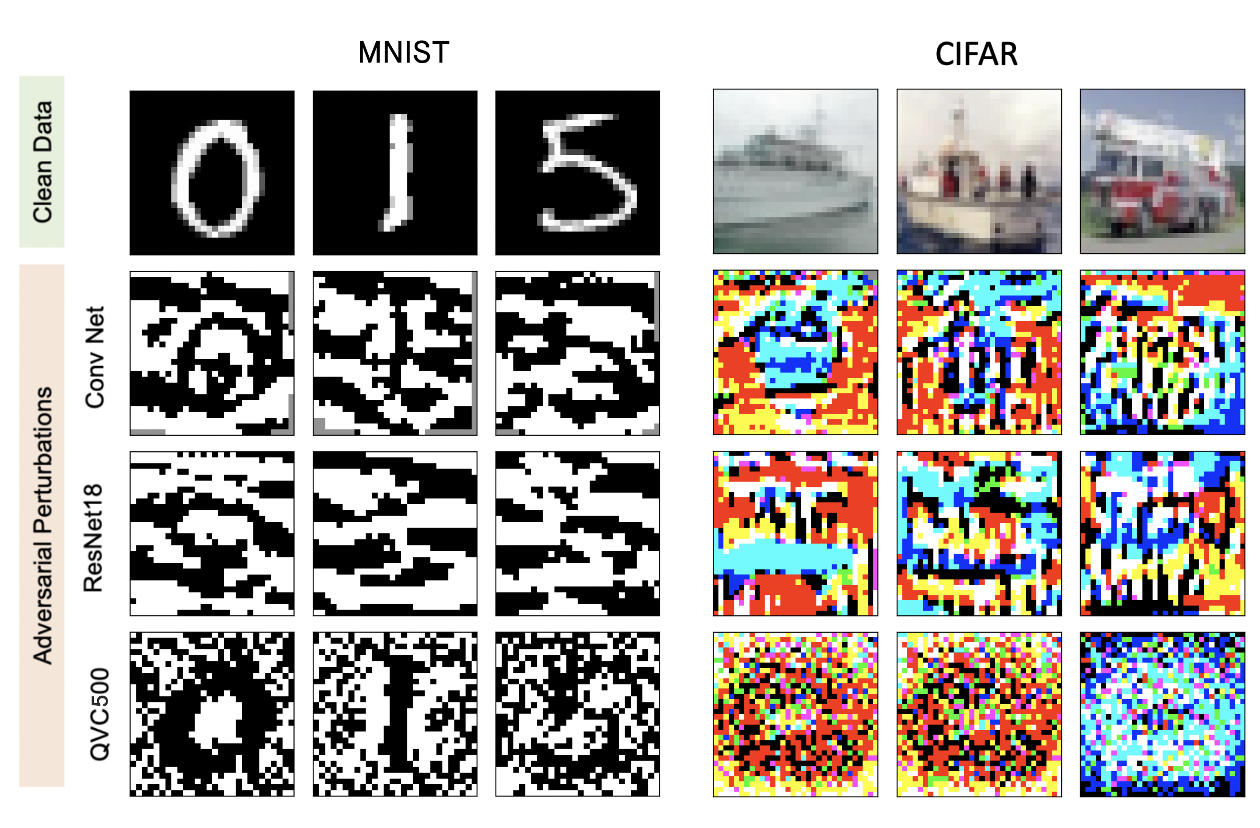}
\label{fig:S9}
   \caption{\label{fig:S9} The adversarial perturbations generated by $\epsilon =0.1$ FGSM attacks on
   the convolutional neural network, ResNet18 and the 500 layer quantum variational classifier (QVC500) are shown for several
   examples from the MNIST and CIFAR databases.
   }
\end{center}
\end{figure*}

\begin{figure*}[ht]
 \begin{center}
 \includegraphics[width=0.7\textwidth]{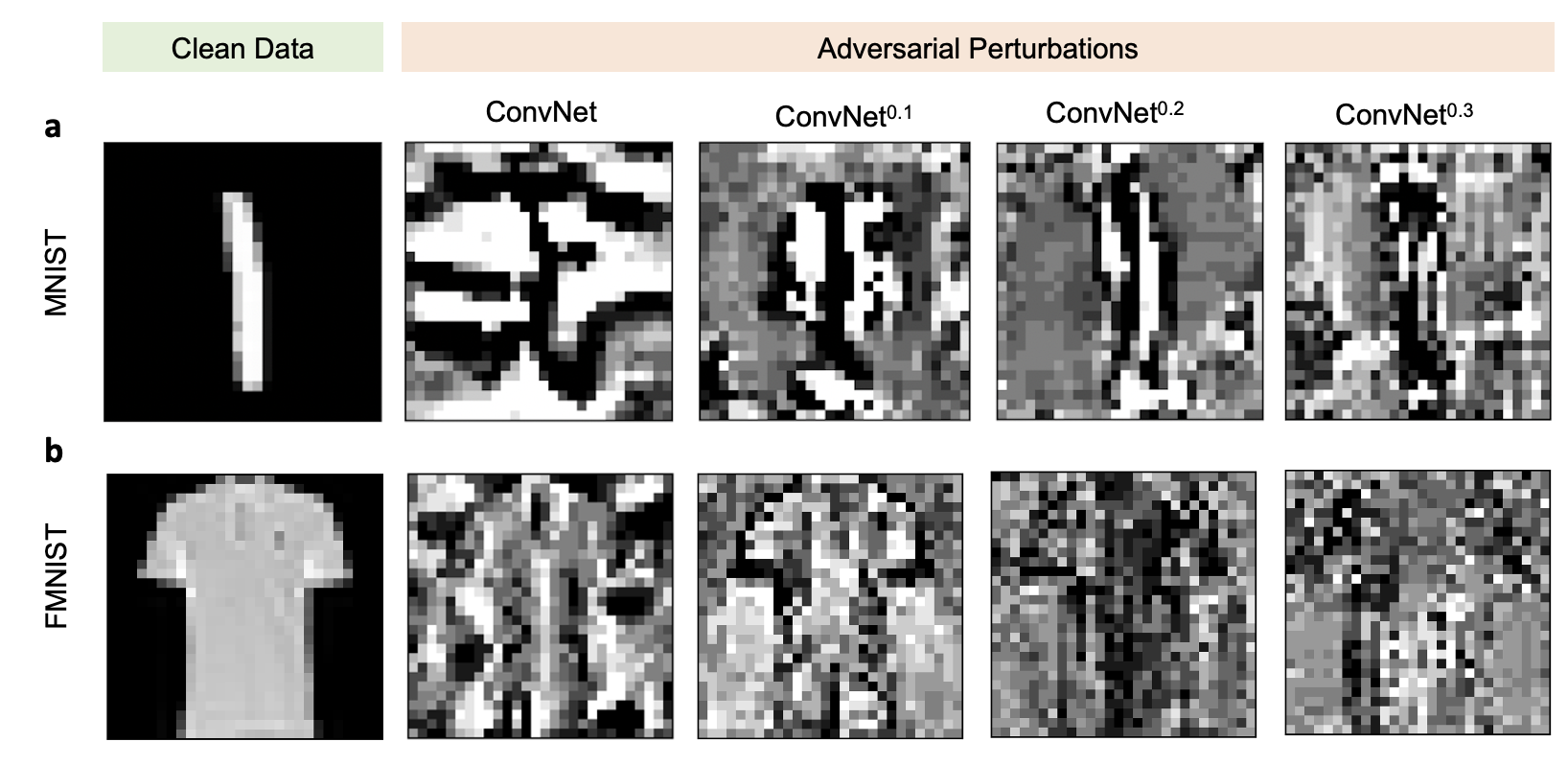}
\label{fig:S10}
    \caption{
    \textbf{a,b.} The adversarial perturbations generated by PGD attacks on ConvNet, ConvNet$^{0.1}$, ConvNet$^{0.2}$ , and ConvNet$^{0.3}$ are shown for sample images from the MNIST and FMNIST datasets. While the perturbation for ConvNet exhibits highly abstract features which are incomprehensible to humans, the perturbations from the adversarial training show hints of systematic features, reminiscence of quantum networks.}
\end{center}
\end{figure*}

\begin{figure*}[ht]
 \begin{center}
 \includegraphics[width=\textwidth]{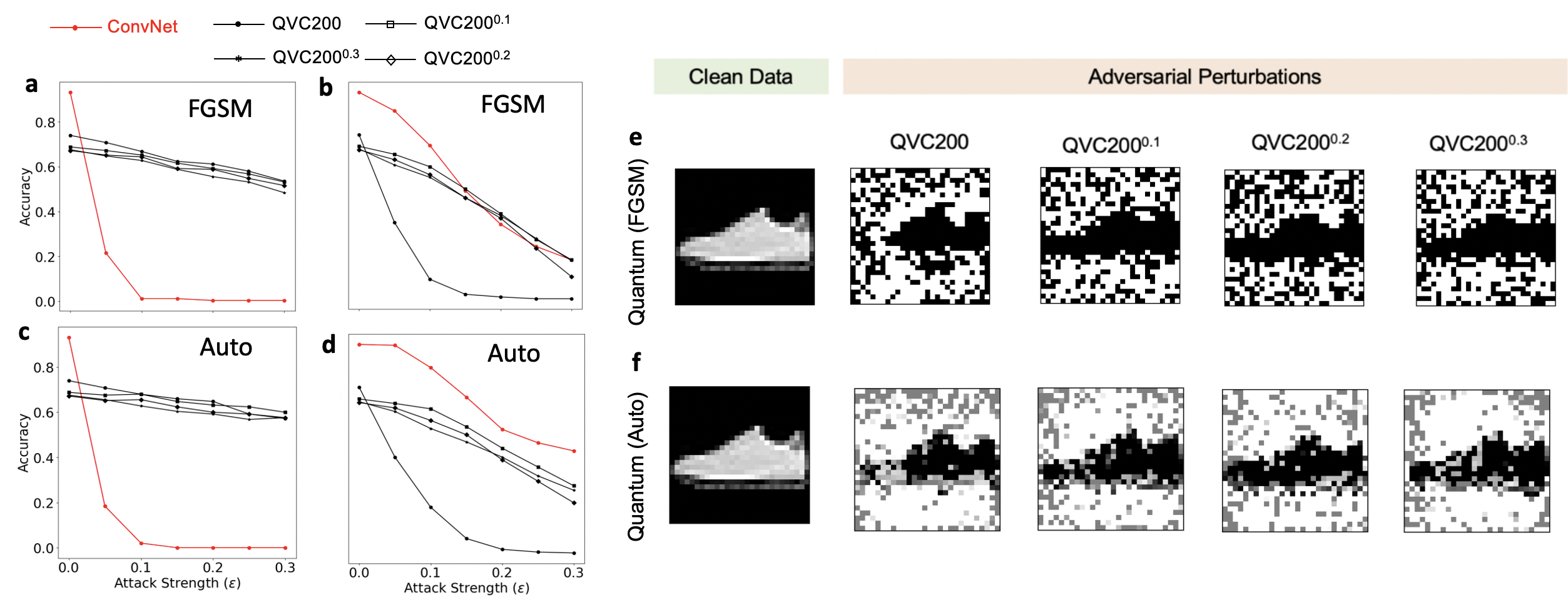}
\label{fig:S11}
    \caption{
   As in Figure 4 of the main text, but using the FGSM and Auto attacks for FMNIST dataset. }
\end{center}
\end{figure*}

\begin{figure*}[ht]
 \begin{center}
 \includegraphics[width=0.5\textwidth]{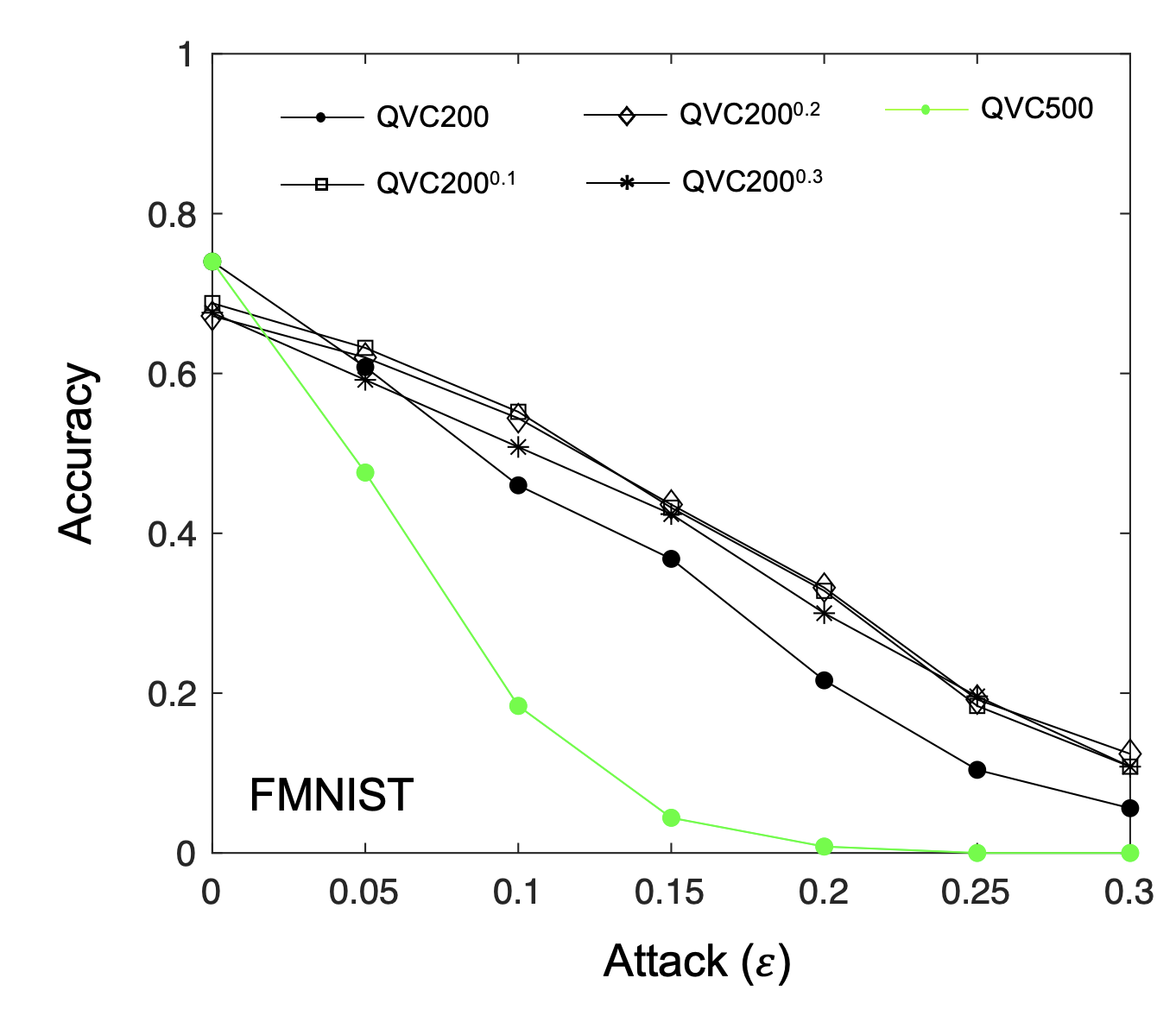}
\label{fig:S12}
\caption{\label{fig:S12} The accuracies achieved by quantum ML networks are plotted as a function of attack strength on a set of 250 adversarially attacked test images from the FMNIST dataset in the cases of white box attacks on the QVC500 network. The attack is applied to QVC200 and adversarially trained QVC200 networks (QVC200$^{0.1}$, QVC200$^{0.2}$, and QVC200$^{0.3}$ where adversarial training is performed with PGD attacks of $(l_{\infty})$ strength 0.1, 0.2 and 0.3 respectively). The adversarial training makes a negligible difference to the accuracy of the quantum network. 
   }
\end{center}
\end{figure*}

\begin{figure*}[ht]
 \begin{center}
 \includegraphics[width=0.45\textwidth]{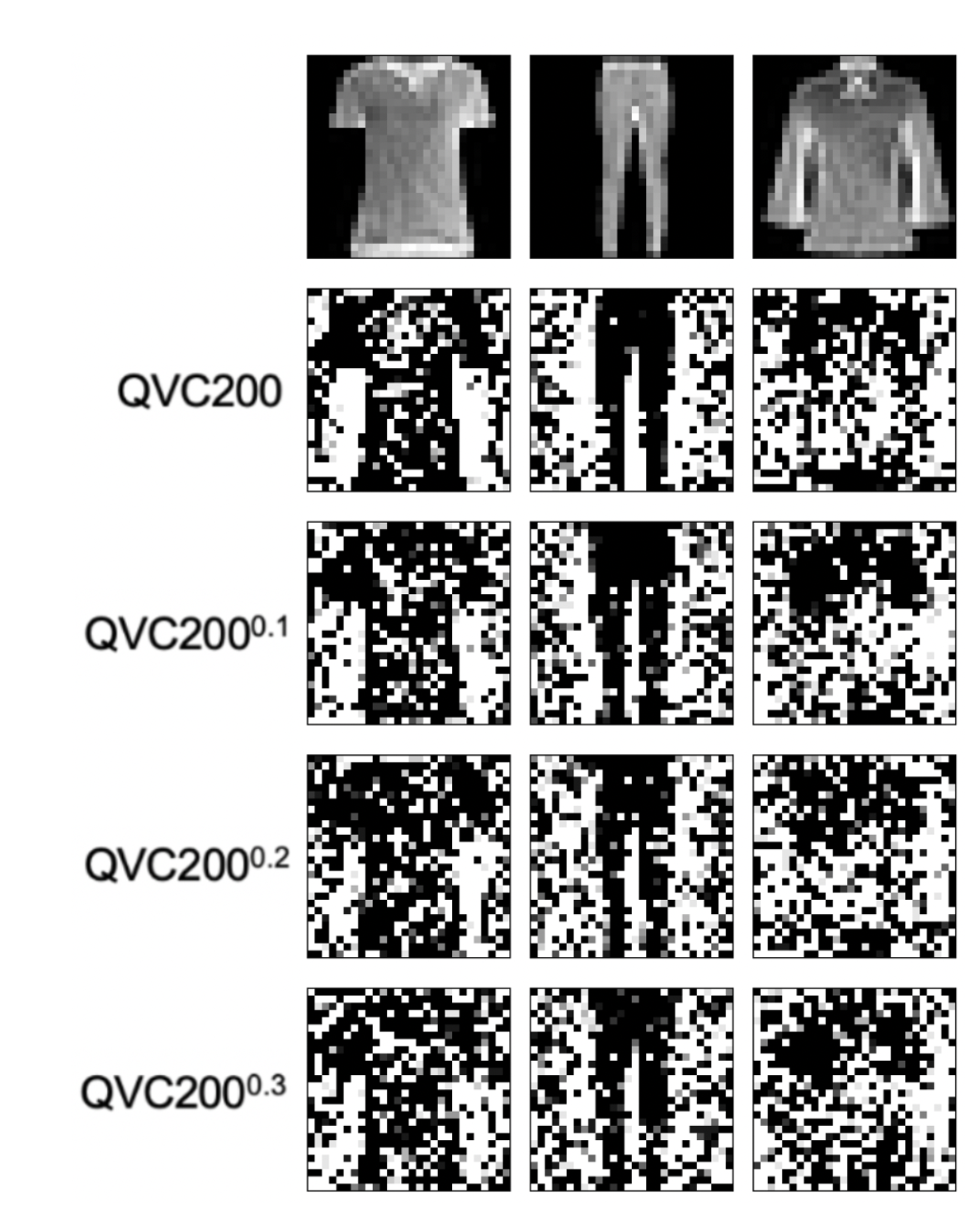}
\label{fig:S13}
\caption{\label{fig:S12} The adversarial perturbations generated by $\epsilon =0.1$ PGD attacks on QVC200 networks, with and without adversarial training, for a few
   examples from the FMNIST dataset. The QVCs display large-scale features whose meaning is often understandable, unlike ConvNet with standard training.
   }
\end{center}
\end{figure*}

\begin{figure*}[ht]
 \begin{center}
 \includegraphics[width=0.5\textwidth]{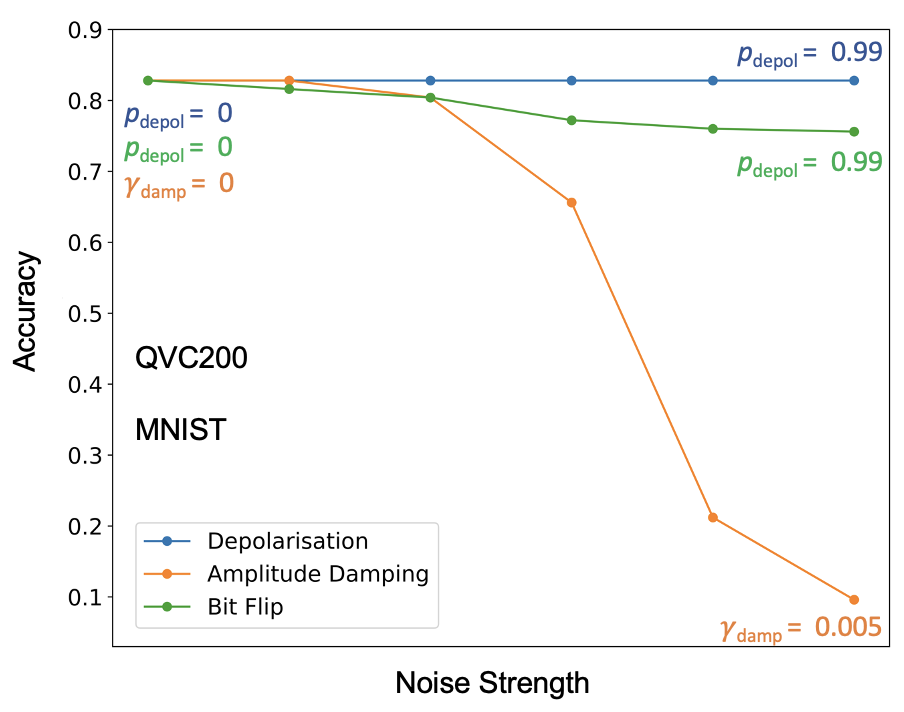}
\label{fig:S14}
    \caption{
   The diagram exhibiting the plots of the learning accuracy of QVC200 network for the MNIST dataset under various noisy simulation environments. The description of noise models is provided in the Methods section. }
\end{center}
\end{figure*}

\begin{figure*}[ht]
 \begin{center}
 \includegraphics[width=0.55\textwidth]{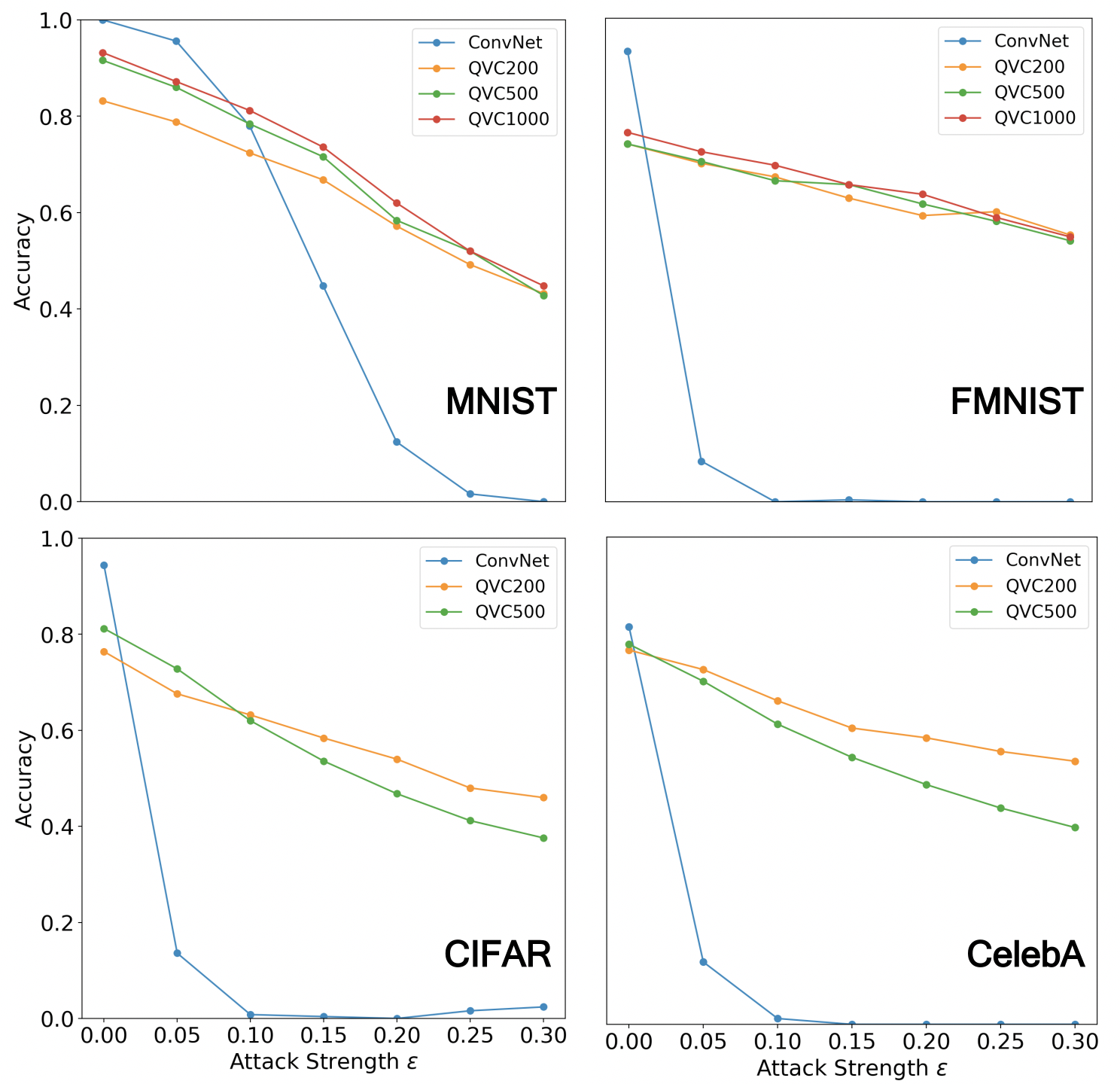}
\label{fig:S15}
    \caption{The accuracies of QVCs of various depths on images generated by a PGD attack on ConvNet. We find that increasing the depth of the quantum networks has a limited effect on both clean accuracy and adversarial robustness. }
\end{center}
\end{figure*}

\end{document}